\documentclass[showpacs, pra, aps, twocolumn,10pt]{revtex4}
\usepackage{mathrsfs}
\usepackage{array}
\usepackage{amsmath}
\usepackage{graphicx}
\usepackage{enumerate}
\usepackage{amstext}
\usepackage{bm}
\usepackage{amsfonts}
\usepackage[utf8]{inputenc}
\usepackage{float}

\begin{document}

\title{Remote preparation of motional Schr\"{o}dinger cat states via dissipatively-driven non-Gaussian mechanical entanglement}

\author{Zunbo Yu}
\affiliation{Department of Physics, Huazhong Normal University, Wuhan 430079, China}
\author{Miaomiao Wei }
\affiliation{Department of Physics, Huazhong Normal University, Wuhan 430079, China}
\author{Huatang Tan}
\email{tht@mail.ccnu.edu.cn}
\affiliation{Department of Physics, Huazhong Normal University, Wuhan 430079, China}
\begin{abstract}
In this paper, we propose a driven-dissipative scheme for generating  non-Gaussian mechanical entangled states and remotely preparing mechanical Schr\"{o}dinger cat states via the entanglement. The system under study consists of a cavity optomechanical setup with two frequency-mismatched mechanical oscillators coupled to a cavity field driven by a bichromatic pump. We show that under proper conditions, an effective Hamiltonian for nondegenerate parametric downconversion involving the two mechanical oscillators and the cavity field can be engineered. We demonstrate analytically and numerically that the cavity dissipation drives the mechanical oscillators into a steady-state pair-coherent state. The no-Gaussianity and nonclassical properties, including Winger negativity, entanglement and quantum steering, of the achieved non-Gaussian mechanical state are investigated in detail. We further show that homodyne detection on one mechanical oscillator enables the remote generation of Schr\"{o}dinger cat states in the other oscillator through the non-Gaussian mechanical entanglement. As we show, this detection can be implemented by transferring the mechanical state to the output field of an auxiliary probe cavity coupled to the target oscillator, followed by homodyne detection on the output field. We also discuss the robustness of the mechanical entangled states and cat states against thermal fluctuations. Our findings establish a feasible approach for the dissipative and remote preparation of mechanical nonclassical states.
\end{abstract}
\maketitle

\section{Introduction}
Cavity optomechanics describes the interaction between the electromagnetic field inside a cavity and a mechanical resonator \cite{mp01,mp02,mp03}. In recent years, significant progress has been achieved in ground-state cooling \cite{mp07, mp08} and the preparation of nonclassical states \cite{mpex1,mpex2,mpex3,ngex3,ngex2} of mechanical oscillators by cavity optomechanics. The preparation of diverse macroscopic quantum states in mechanical oscillators holds significant importance for testing fundamental assumptions of quantum mechanics and advancing quantum technologies \cite{mp04,mp05,mp06}.
Quantum reservoir engineering, which utilizes engineered dissipation to achieve desirable quantum states,  has been employed to achieve squeezed and entangled mechanical states \cite{mpt,drex1,drex2,drex3}.
However, by the linearized cavity optomechanics only can Gaussian mechanical states be obtained.
In contrast, non-Gaussian quantum states, which may exhibit negative Wigner functions and shows genuine quantum nonclassicality \cite{ngs1}, proves to be unique advantages in the applications of quantum technologies such as quantum computation\cite{ngs01,ngs02,ngs2}, quantum communication\cite{mp10,mp11,mp12}, and quantum metrology\cite{ngs03,ngs04,ngs05}. Consequently, the preparation of various intriguing non-Gaussian states in mechanical oscillators has been explored by leveraging the nonlinear optomechanical coupling or measurements \cite{ngth1,ngth2,ngex1,ngex4,ngex5,ngs06}. For example, the non-Gaussian mechanical states have been experimentally generated via phonon addition or subtraction for a initially thermalized mechanical oscillator\cite{ngm1,ngm2}. Nonlinear engineered dissipation such as two phonon absorption has been considered to prepare mechanical quantum superpositions \cite{mpt4}. These locally generated nonclassical states will suffer from transmission losses when applied to realistic quantum information protocols.


Remote state preparation (RSP) is a novel quantum state transmission protocol just based on shared quantum entanglement\cite{rsp1}.
As a method of quantum state transmission, RSP offers several advantages compared to quantum teleportation \cite{rsp2}, including reduced consumption of classical communication resources and the elimination of the need for Bell-state measurements \cite{rsp3}. Additionally, when compared to the direct transmission of quantum states, RSP provides enhanced security \cite{rsp4}.
Research has demonstrated that RSP plays a significant role in understanding the underlying logic of certain quantum computing and quantum communication schemes \cite{rsp5,rsp6,rsp7}.
Very recently, the remote preparation of optical non-Gaussian states and magnon Schr\"{o}dinger cat states has been experimentally realized \cite{hesu,mp16}.

In this paper, we propose a driven-dissipative scheme for generating steady non-Gaussian entangled states of mechanical oscillators and remotely preparing  mechanical Schr\"{o}dinger cat states through the non-Gaussian entanglement. The system under study consists of a cavity optomechanical setup in which two frequency-mismatched mechanical oscillators are coupled to a cavity field driven by a bichromatic pump. We demonstrate explicitly that, under proper conditions, an effective Hamiltonian for nondegenerate downconversion involving the two mechanical oscillators and the cavity field can be engineered. We analytically derive its steady-state solution as a pair-coherent state (PCS) \cite{pcsp1,pcsp5} of the mechanical oscillators, which exhibits non-Gaussian nonclassicalties, as we reveal in detail in the context. We note that the PCS of micrwave field have already been experimentally realized very recently in a circuit-QED system \cite{pcsp4}. It has also been proven that the PCS has potential application e.g. in quantum computation \cite{pcsp6}.
We show that homodyne detection on one mechanical oscillator can generate Schr\"{o}dinger cat states in another mechanical oscillator via the mechanical non-Gaussian entanglement, and we also propose a scheme to implement the detection by transferring the mechanical state to the output field of an auxiliary probe cavity coupled to the target oscillator and then homodyning the output field, realizing the remote state preparation. We also discuss the effects of thermal fluctuations on the obtained mechanical states. Our findings establish a feasible approach for the dissipative and remote preparation of mechanical nonclassical states.

The paper is organized as follows.
In Sec.II, the effective Hamiltonian and master equation for the present system is derived. The steady-state solution was derived analytically and verified numerically.
In Sec. III,
the fidelity, non-Gaussianity, and nonclassicalities of the achieved mechanical entangled state are investigated in detail.
In Section IV,
the remote preparation of mechanical Schr\"{o}dinger cat state via the non-Gaussian mechanical entanglement is investigated.
In Sec.V, the effect of thermal fluctuations on the present scheme is estimated.
In Sec.VI., the conclusion is given.

\section{Dissipatively-driven non-Gaussian mechanical entangled states}
\begin{figure}
\centerline{\scalebox{0.20}{\includegraphics{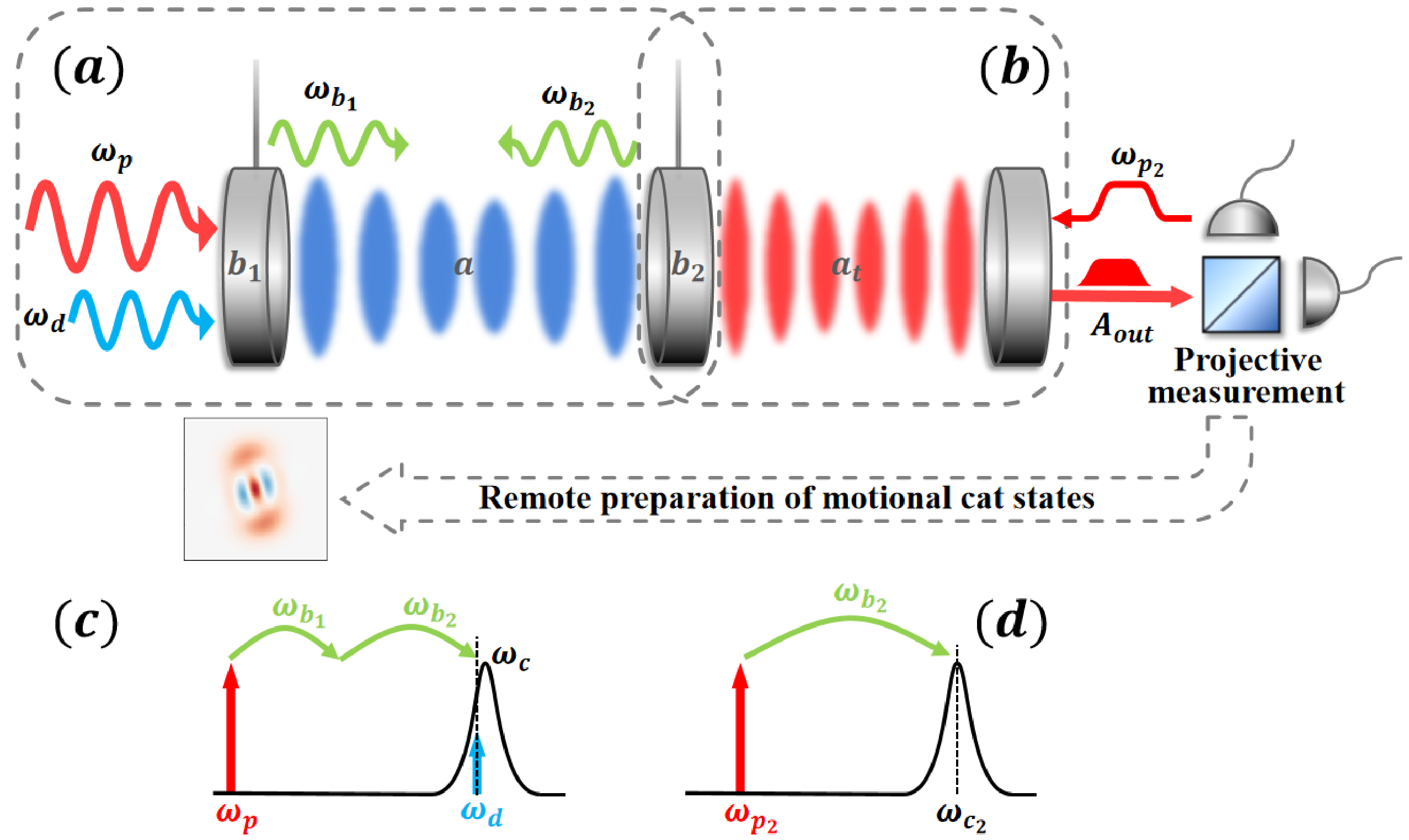}}}
\vspace{0cm}
\caption{(a) Schematic plot of a cavity optomechanical system in which two mechanical oscillators are dispersively coupled to a bichromatically-driven cavity. (b) The state transfer from a mechanical oscillator to a probe cavity driven by a plused laser to achieve homodyne detection on the mechanical oscillator for remote-state preparation. (c) and (d) The frequencies of the mechanical oscillators ($\omega_{b_1}$, $\omega_{b_2}$), the optomechanical and probe cavities ($\omega_c$, $\omega_{c_2}$), and the drive fields ($\omega_p$, $\omega_d$ and $\omega_{p_2}$).}
\label{sys1}
\end{figure}

As shown in Fig.\ref{sys1}~(a), we consider a cavity optomechanical system in which the cavity field dispersively interacts with two mechanical oscillators and is simultaneously driven by both a strong and a weak laser. The interaction of the system can be described by the Hamiltonian
\begin{align}
\hat H_1=&\omega_{c}\hat{a}^{\dagger}\hat{a}+\sum_{j=1}^{2}\omega_{b_j}\hat{b}_j^{\dagger}\hat{b}_j+g_j\hat{a}^{\dagger}\hat{a}\big(\hat{b}_j+\hat{b}_j^{\dagger}\big)\nonumber\\
&+\left(\varepsilon_{p}\hat{a}^{\dagger}e^{-i\omega_{p}t}+\varepsilon_{d}\hat{a}^{\dagger}e^{-i\omega_{d}t}+h.c.\right),
\end{align}
where the annihilation operator $\hat{a}~(\hat{b}_j)$ denotes the cavity field (mechanical oscillator) with resonant frequency $\omega_{c}$ ($\omega_{b_j}$), $g_j$ are the single-photon optomechanical coupling rates, and $\omega_{p}$~($\varepsilon_p$) and $\omega_{d}$ ($\varepsilon_{d}$) are frequencies~(amplitudes) of the drive fields with the relations
\begin{align}
\omega_d=\omega_p+\omega_{b_1}+\omega_{b_2}, ~~\omega_d\approx \omega_c,~~\varepsilon_p\gg\varepsilon_d,
\label{relt}
\end{align}
as depicted in Fig.\ref{sys1}~(c).

Including the cavity dissipation and mechanical damping, the density operator $\hat \rho_{ab}$ of the system is satisfied by the following master equation
\begin{align}
\frac{d}{dt}{\hat \rho_{ab}} =& -i[\hat H_1,\hat \rho_{ab}]+\gamma_{a}\mathcal L[\hat{a}]\hat \rho_{ab}\nonumber\\
&+\sum_j\gamma_{b_j}(\bar{n}_{b_j}+1)\mathcal L[\hat{b}_j]\hat \rho_{ab}
+\gamma_{b_j}\bar{n}_{b_j}\mathcal L[\hat{b}_j^\dagger]\hat \rho_{ab},
\label{ms_o}
\end{align}
where $\gamma_a$ is the dissipation rate of the cavity field in a vacuum environment and $\gamma_{b_j}$ are the mechanical damping rates in thermal environments with the mean thermal photon numbers $\bar{n}_{b_j}$.

By using the drive frequency $\omega_{p}$ as the rotating frame and displacing the cavity by the amplitude $\alpha = \frac{\varepsilon_{p}}{i\gamma_a - \Delta}$, where $\Delta=\omega_{c}-\omega_{p}$, applying the unitary transformation $e^{\alpha^{*}\hat{a}-\alpha \hat{a}^{\dagger}}e^{i\omega_{p}\hat{a}^{\dagger}\hat{a}t}$ to the Hamiltonian $\hat H_1$, the resulting Hamiltonian is then given by
\begin{align}
\hat H_2=&\Delta \hat{a}^{\dagger }\hat{a}+\sum_j\Big[\omega_{b_j}\hat{b}_j^{\dagger }\hat{b}_j+
g_{j}\hat{a}^{\dagger }\hat{a}(\hat{b}_j+\hat{b}_j^{\dagger})\nonumber\\
&+(\alpha^{*}\hat{a}+\alpha \hat{a}^{\dagger}+|\alpha|^{2})(g_{j}\hat{b}_j+g_{j}\hat{b}_j^{\dagger})\Big]\nonumber\\
&+\varepsilon_{d}\hat{a}^{\dagger}e^{i\Delta_pt}
+\varepsilon_{d}^{*}\hat{a}e^{-i\Delta_p t},
\end{align}
where $\Delta_p = \omega_p-\omega_d$. The third and fourth parts in the Hamiltonian respectively present the nonlinear and linearized optomechanical interactions. Generally, in linearized cavity optomechanics, the nonlinear interactions are neglected for very weak single-photon optomechancal coupling, i.e., $r_j\equiv g_j/\omega_j\ll1$. In contrast, here we retain the nonlinear terms by assuming moderate single optomechanical coupling (e.g., $r_j\sim0.1$), which may be reached experimentally \cite{exp1,exp2,exp3,nbe1,nbe2}.

We proceed to apply two successive Schrieffer-Wolff transformations, described by the generator $\hat S^{\dagger}=e^{r_2 \hat{a}^{\dagger }\hat{a}(\hat{b}_2^{\dagger }-\hat{b}_2)}e^{r_1\hat{a}^{\dagger }\hat{a}(\hat{b}_1^{\dagger }-\hat{b}_1)}$, to the Hamiltonian $\hat H_2$, resulting in the Hamiltonian
\begin{align}
\hat H_3=&\widetilde{\Delta} \hat{a}^{\dagger}\hat{a}+\omega_{b_1}\hat{b}_1^{\dagger}\hat{b}_1+\omega_{b_2}\hat{b}_2^{\dagger}\hat{b}_2
+g_0
\hat{a}^{\dagger}\hat{a}\hat{a}^{\dagger}\hat{a}\nonumber\\
&+(\alpha^{*}\hat{ \widetilde{a}}+\alpha \hat{ \widetilde{a}}^{\dagger})(g_1 \hat{b}_1+g_1 \hat{b}_1^{\dagger}+g_{2}\hat{b}_2+g_{2}\hat{b}_2^{\dagger })\nonumber\\
&+|\alpha|^2 (g_1 \hat{b}_1+g_1 \hat{b}_1^{\dagger}+g_{2}\hat{b}_2+g_{2}\hat{b}_2^{\dagger })
\nonumber\\
&+2g_0\hat{a}^{\dagger}\hat{a}(\alpha^{*}\hat{\widetilde{a}}+\alpha \hat{\widetilde{a}}^{\dagger})+\varepsilon_{d}\hat{ \widetilde{a}}^{\dagger}e^{i\Delta_p t}+\varepsilon_{d}^{*}\hat{ \widetilde{a}}e^{-i\Delta_p t},
\end{align}
where the detuning $\widetilde{\Delta}=\left(\Delta+2|\alpha|^2g_0\right)\nonumber$ and the coupling $g_0 =- \left(r_1g_1+r_2g_2\right)$.  The operator $\hat{ \widetilde{a}}^{\dagger}
=S^{\dagger}\hat{a}^{\dagger}S =\hat{a}^{\dagger}e^{r_1(\hat{b}_1^{\dagger }-\hat{b}_1)}e^{r_2(\hat{b}_2^{\dagger }-\hat{b}_2)}$,
which can be approximated, for $r_{j}\gg r_{j} r_{j'}$, as
\begin{align}
\hat{ \widetilde{a}}^{\dagger}
=& \hat{a}^{\dagger }\Big[1+\frac{r_1}{1!}(\hat{b}_1^{\dagger}-\hat{b}_1)+\frac{(r_1)^2}{2!}(\hat{b}_1^{\dagger}-\hat{b}_1)^{2}+\cdots \Big]\nonumber\\
&\times\Big[1+\frac{r_2}{1!}(\hat{b}_2^{\dagger}-\hat{b}_2)+\frac{(r_2)^2}{2!}(\hat{b}_2^{+}-\hat{b}_2)^{2}+\cdots \Big]\nonumber\\
\approx& \hat{a}^{\dagger}\Big[1+r_1(\hat{b}_1^{\dagger}-\hat{b}_1)
+r_2(\hat{b}_2^{\dagger}-\hat{b}_2)\Big].
\label{efa}
\end{align}
Substituting the expression of $\hat{ \widetilde{a}}^{\dagger}$ into the Hamiltonian $\hat H_3$ and applying the unitary transformation $e^{i\hat{\widetilde{ H_0}}t}$, where $
\hat{\widetilde{H_0}}=\widetilde{\Delta}\hat{a}^{\dagger}\hat{a}+\omega_{b_1}\hat{b}_1^{\dagger}\hat{b}_1+\omega_{b_2}\hat{b}_2^{\dagger}\hat{b}_2 $, it is changed, under the condition $\widetilde{\Delta}=-\Delta_p=\omega_{b_1}+\omega_{b_2}$ (as described in Eq.\ref{relt}), into (see the Appendix)
\begin{align}
\hat H_4
\approx g_0
\hat{a}^{\dagger}\hat{a}\hat{a}^{\dagger}\hat{a}
+g\hat{a}^{\dagger}\hat{b}_1 \hat{b}_2 + g^{*}\hat{a} \hat{b}_1^{\dagger} \hat{b}_2^{\dagger} + \varepsilon_{d}\hat{a}^{\dagger} + \varepsilon_{d}^{*}\hat{a},
\label{h_a}
\end{align}
where $g = -\alpha \left(r_1g_2+r_2g_1\right)$. The above Hamiltonian is obtained also under the rotating wave approximation requiring the following conditions
\begin{align}
\label{eqcon}
\{\omega_{b_j},&|\omega_{b_1}-\omega_{b_2}|,\widetilde{\Delta},
|\widetilde{\Delta}-\omega_{b_j}|\}\gg\\ \nonumber
&\{r_j\varepsilon_{d},g_j|\alpha|^2,
 \alpha \left(r_1g_2+r_2g_1\right),
\alpha \left(r_1g_1+r_2g_2\right)\}.
\end{align}
In Eq.(\ref{h_a}), the first term is the effective cavity Kerr nonlinearity, the second (third) term describes the hybrid four-wave mixing process involving the cavity photons, strong-drive photons and phonons, in which a drive photon and two nondegenerate phonons are simultaneously annihilated and then a cavity photon ($\omega_c\approx\omega_p+\sum_j\omega_{b_j}$) is created (vice versa). Since the strong-drive photons are treated classically, the cavity photons and phonons form the effective non-degenerate parametric downconversion with the quantized pump. This process  may establish non-Gaussian quantum correlations between the two mechanical oscillators, under the cavity driving on resonance described by the last two terms.

In the transformed picture, the master equation (\ref{ms_b}) is approximately reduced to (see the Appendix)
\begin{align}
\frac{d}{dt} \hat \rho_{ab}
&\approx -i\left[ \hat H_4 , \hat\rho_{ab} \right]+ \gamma_{a}\mathcal L[\hat{a}]\hat\rho_{ab}\nonumber\\
&+\sum_j\gamma_{b_j}(\bar{n}_{b_j}+1)\mathcal L[\hat{b}_j]\hat\rho_{ab}
+\gamma_{b_j}\bar{n}_{b_j}\mathcal L[\hat{b}_j^\dagger]\hat\rho_{ab}.
\label{ms_b}
\end{align}
It can be inferred from the above master equation that, in the absence of the mechanical damping ($\gamma_{b_j}=0$), the system of the cavity and mechanical oscillators is dissipated by the cavity dissipation into a steady dark state, i.e., $\hat \rho_{ab}^{\rm ss}=|\zeta\rangle_b|0\rangle_a \langle 0|_b\langle \zeta|$, where the cavity field is in vacuum and the mechanical state $\left | \zeta \right \rangle_b$  satisfies the equation $(g\hat{b}_1\hat{b}_2+\varepsilon _d)\left | \zeta \right \rangle_b=0$. This shows that the two mechanical oscillators are dissipatively driven into the eigenstate of the operator $\hat{b}_1\hat{b}_2$, i.e.,
\begin{align}
 \hat{b}_1\hat{b}_2\left | \zeta \right \rangle_b=\zeta \left | \zeta \right \rangle_b,
 \end{align}
with the eigenvalue $\zeta = -\varepsilon _d/g$. This state is a PCS \cite{pcsp1}, which is analogous to a coherent state as the eigenstate of a bosonic annihilation operator.
For simplicity, here we assume that the initial phonon number difference between the two mechanical oscillators \(\hat{b}_1^{\dagger}\hat{b}_1 - \hat{b}_2^{\dagger}\hat{b}_2 = 0\) (e.g., both start from vacua), and the explicit expression for the PCS in the Fock space can be obtained as
\begin{align}
\left | \zeta \right \rangle_b & \propto\sum_{n=0 }^{\infty }
\frac{\zeta ^{n} }{n! }   |  n   \rangle_1| n \rangle_2.
\label{pcs}
\end{align}
Similar to a two-mode squeezed vacuum, the PCS is also a superposition only of states in which the two modes contain the same number of phonons. However,  the probability amplitudes of the PCS differ from those of the two-mode squeezed vacuum, which has Gaussian-type quasiprobability distribution in phase space. As a result, the PCS is a non-Gaussian state, whose non-Gaussian characteristics will be studied later.
\begin{figure}
\centerline{\hspace{-0.25cm}\scalebox{0.114}{\includegraphics{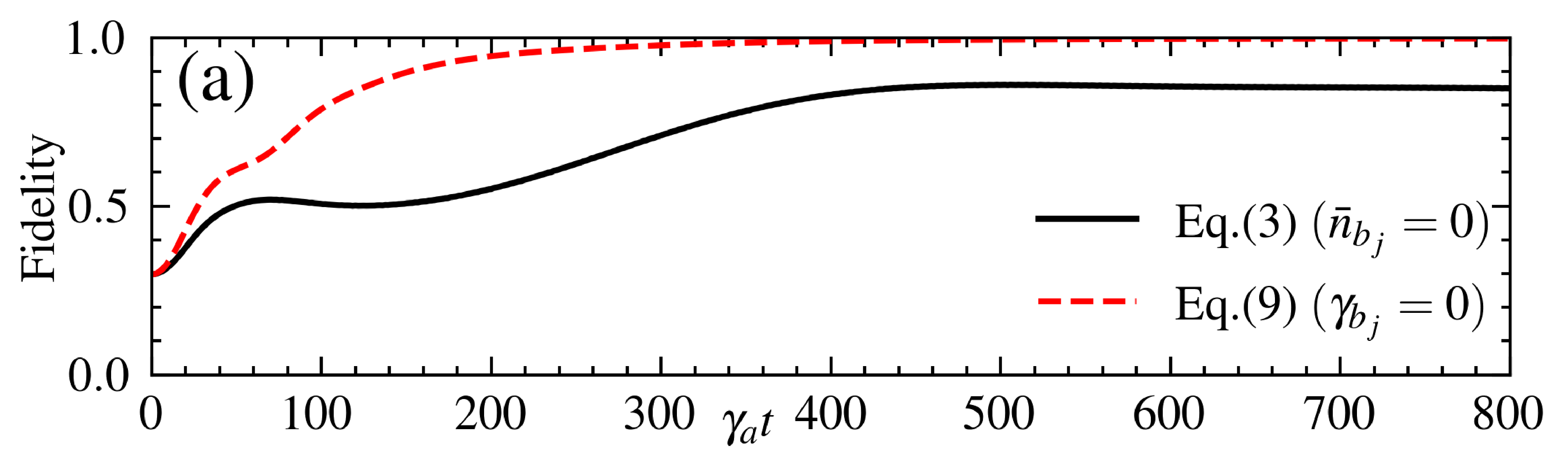}}}
\vspace{-0.16cm}
\centerline{\scalebox{0.085}{~~\includegraphics{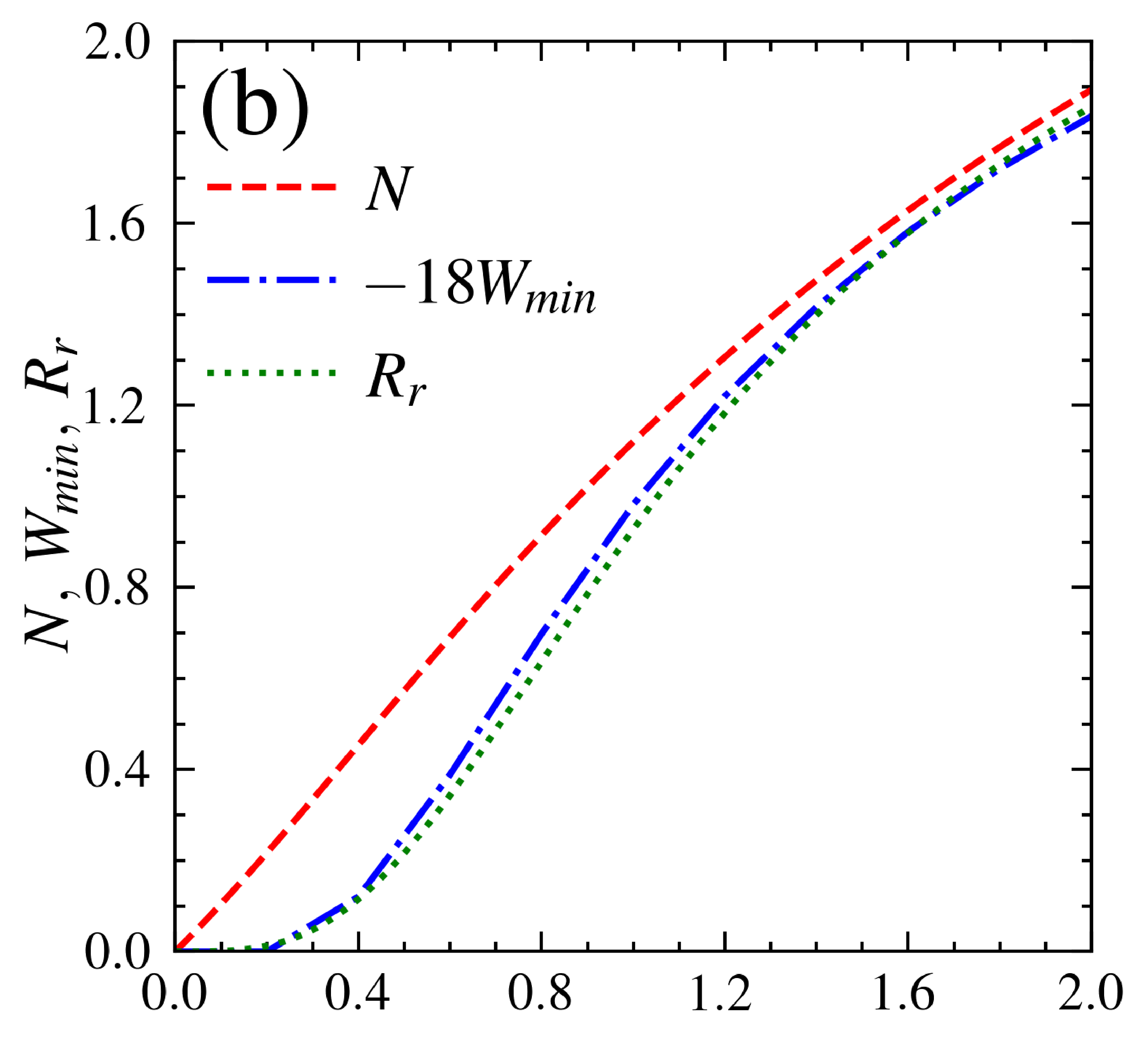}\includegraphics{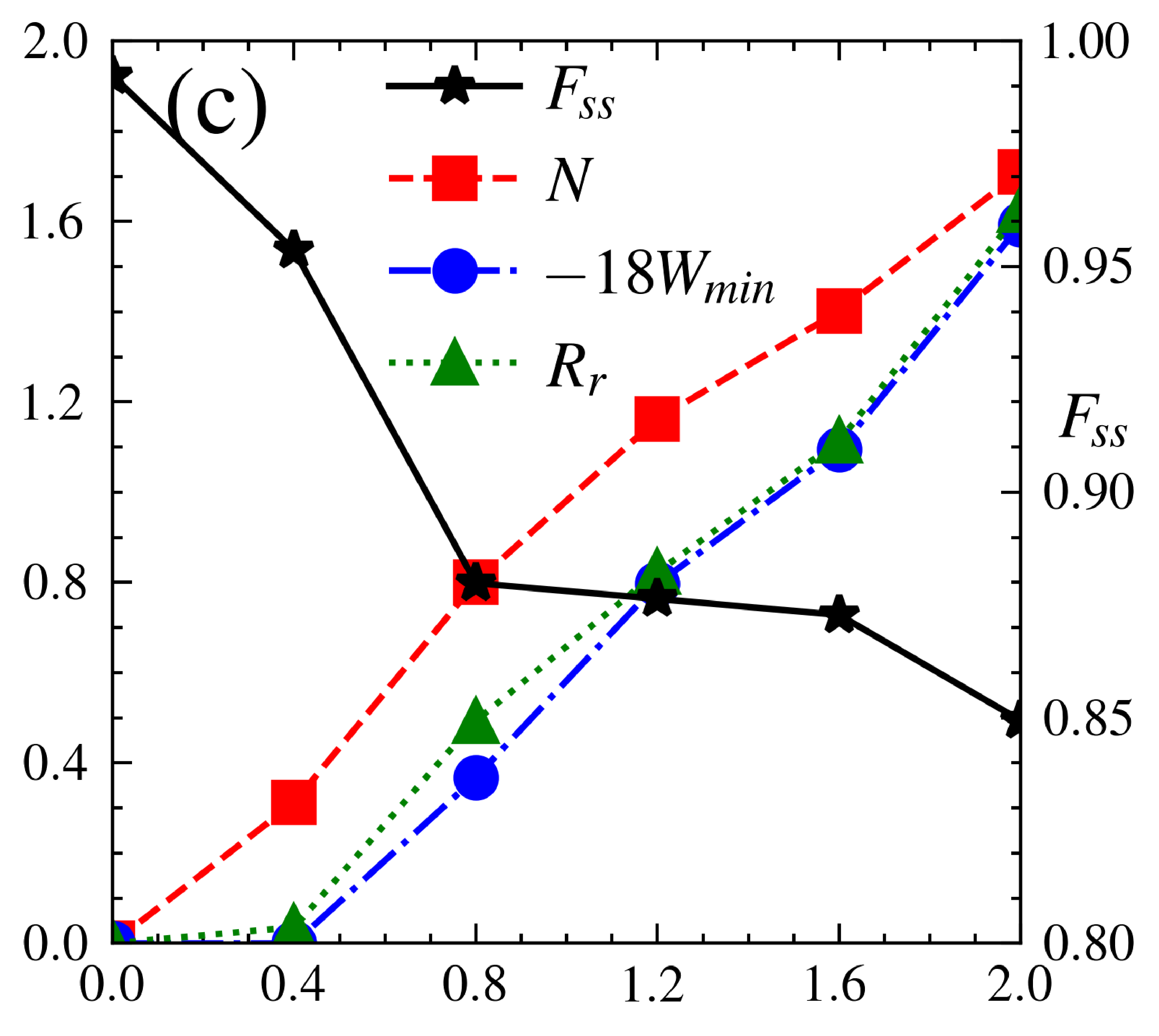}}}
\vspace{-0.15cm}
\centerline{\hspace{0.1cm}\scalebox{0.084}{~~\includegraphics{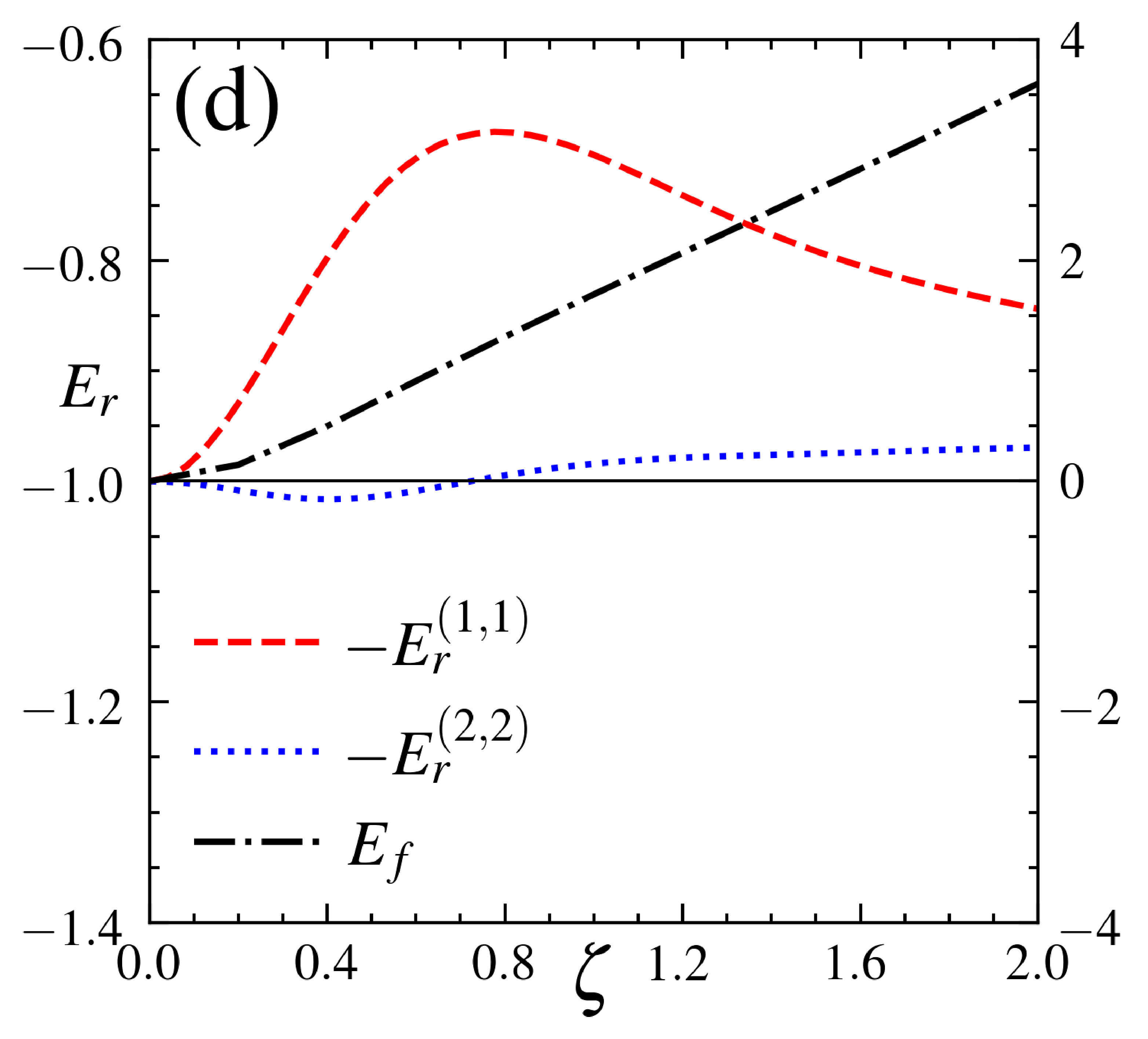}\hspace{-1.4cm}\includegraphics{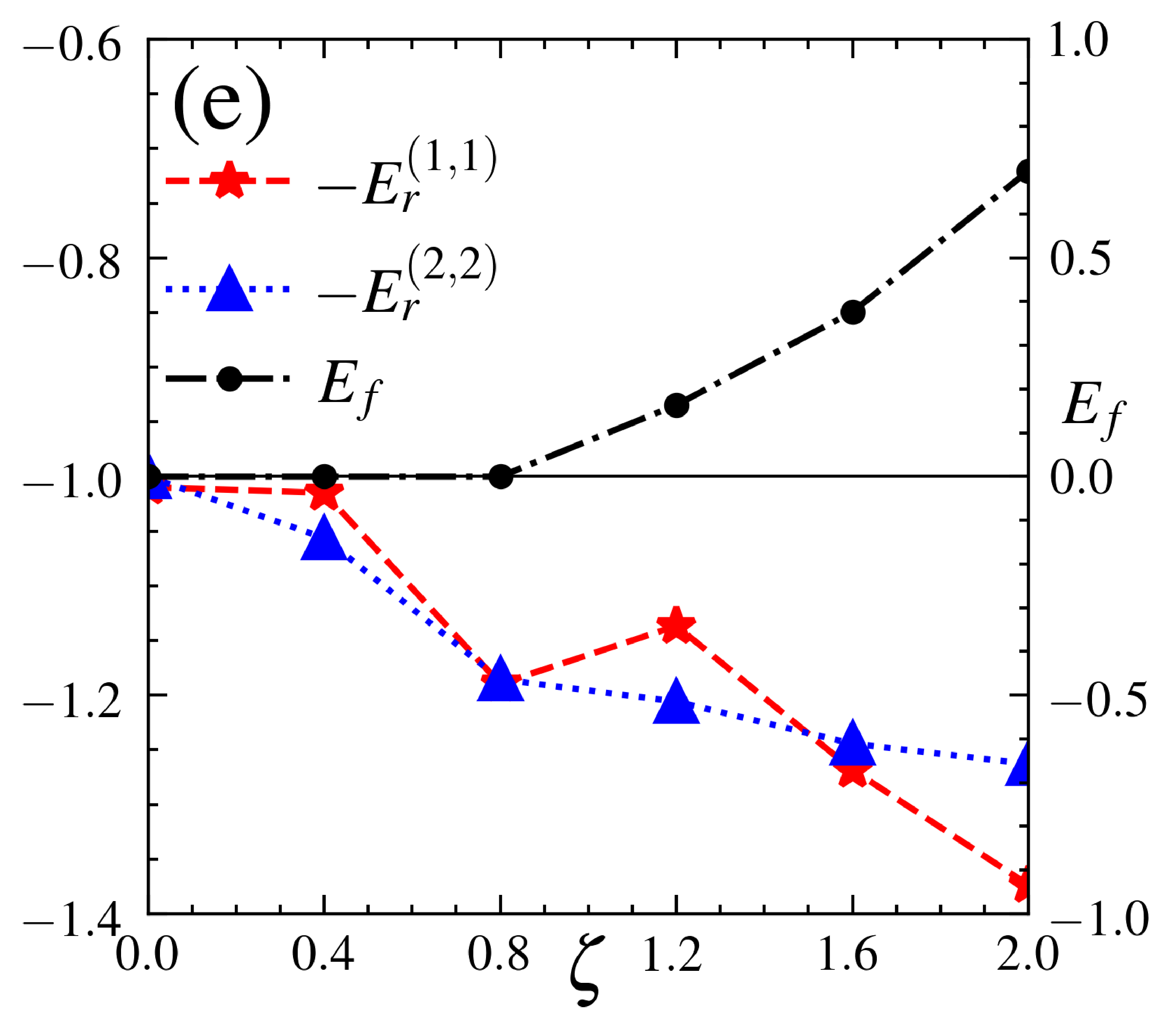}}}
\caption{(a) The time evolution of the fidelity $F$ of the mechanical state $\hat \rho_b$ to the pure PCS $|\zeta=2\rangle_b$ via numerically solving the full master equation (\ref{ms_o}) and the approximate equation (\ref{ms_b}).
(b) and (c) The dependences of the mechanical entanglement $\mathcal N$, non-gaussianity $R_r$, and minimal Wigner negativity $W_{min}^b$ of the pure PCS $|\zeta\rangle$ (b) and the steady mechanical state $\hat \rho_b^{\rm ss}$ (c) on the functions of the parameter $\zeta$. The fidelity $F_{\rm ss}$ of the mechanical state $\hat \rho_b^{\rm ss}$ to the PCS is also plotted in (c).
(d) and (e) The mechanical quantum steering, witnessed by $E_r$ and $E_f$, of the pure PCS (d) and the steady state $\hat \rho_b^{\rm ss}$ (e), as the functions of the parameter $\zeta$.}
\label{fig_sprop}
\end{figure}

In Fig.\ref{fig_sprop}~(a), the time evolution of the fidelity $F$ of the mechanical state $\hat \rho_b\equiv Tr_a(\hat \rho_{ab})$ to the pure mechanical PCS state $|\zeta=2\rangle_b$ is plotted by numerically solving the approximate master equation (\ref{ms_b}) and the full master equation (\ref{ms_o}) after performing the unitary transformation $\hat U^{\dagger}=e^{i\widetilde{H_0}t}e^{r_2 \hat{a}^{\dagger }\hat{a}(\hat{b}_2^{\dagger }-\hat{b}_2)}e^{r_1\hat{a}^{\dagger }\hat{a}(\hat{b}_1^{\dagger }-\hat{b}_1)}e^{\alpha^{*}\hat{a}-\alpha \hat{a}^{\dagger}}e^{i\omega_{p}\hat{a}^{\dagger}\hat{a}t}$.
The parameters are given by $\omega_{b_2}$=1.5$\omega_{b_1}$, $g_1$=0.045$\omega_{b_1}$, $g_2$=0.055$\omega_{b_1}$, $\varepsilon_p$=$1.58\omega_{b_1}$, $\varepsilon_d=-5.218\times10^{-3}\omega_{b_1}$, $\Delta$=$2.5032\omega_{b_1}$, $\Delta_p=-2.5\omega_{b_1}$, $\gamma_a=2.5\times10^{-3}\omega_{b_1}$,  $\gamma_{b_1}=10^{-6}\omega_{b_1}$, $\gamma_{b_2}=10^{-6}\omega_{b_2}$, and $\bar{n}_{b_j}=0 $. The initial state of the cavity and two mechanical oscillators is considered to be vacuum.
It is shown from Fig.\ref{fig_sprop}~(a) that in the long-time limit, the fidelity $F=1$, as revealed by the master equation (\ref{ms_b}) with $\gamma_{b_j}=0$. This shows that the mechanical oscillators are indeed dissiatively into the PCS with the Hamiltonian of Eq.(\ref{eqcon}).
Even with the full master equation (\ref{ms_o}) that includes mechanical damping, under the condition $\bar {n}_{b_j}=0$, the fidelity remains stable, with the long-time value $F_{\rm ss}\approx0.85$, confirming the validness of the approximation used to derive Eq.(\ref{ms_b}).
As shown in Fig.\ref{fig_sprop} (c), where the fidelity of the steady state $\hat \rho_b^{\rm ss}$ is plotted as a function of the parameter $\zeta$ by solving the full master equation (\ref{ms_o}) in the presence of the mechanical damping, the fidelity decreases as $\zeta$ increases. Nevertheless, the high fidelity in the steady-state regime can still be achieved, as we still have $F_{\rm ss}\approx 0.85$ for the state $|\zeta=2\rangle_b$. Therefore, within the present scheme, the two mechanical oscillators are indeed prepared into a steady mixed state $\hat\rho_{b}^{\rm ss}$, which is highly approximated to the PCS state.

\section{Non-Gaussian mechanical entanglement and steering}
In the section, we intend to study the nonclassical features of the achieved mechanical PCS state $\hat \rho_b^{\rm ss}$. Before doing it, let us discuss the non-Gaussianity of the mechanical state, which can be characterized with the quantum relative entropy between $\hat \rho_b^{\rm ss}$ and its Gaussian reference state $\hat\rho_r$ \cite{qre}, i.e.,
\begin{align}
R_r = R\left ( \hat\rho_b^{\rm ss} \right ) -R \left ( \hat\rho_r \right ),
\end{align}
where $R \left ( \hat \rho \right )=-Tr\left[ \hat \rho \ln{(\hat \rho)}\right]$ is the von Neumann entropy of a state $\hat \rho$. The Gaussian reference state is determined merely by its covariance matrix which is the same as that of the non-Gaussian state $\hat \rho_b^{\rm ss}$. The quantum relative entropy $R_r$ is plotted in Fig.\ref{fig_sprop} (b) and (c) respectively for the pure and approximate mechanical PCSs. It is shown that $R_r$ increases as the parameter $\zeta$ increases, indicating the stronger non-Gaussianity of the steady mechanical state with larger $\zeta$. For the same value of $\zeta$, the non-Gaussianity of the pure mechanical PCS is stronger than the corresponding approximate mechanical state. Taking $\zeta=2$ as an example, $R_r\approx 1.86$ for the pure PCS, however, it drops to 1.63 when the mechanical damping and counter-rotating terms  are included.

As we know, a non-Gaussian state may exhibit negative Wigner function which indicates genuine nonclassicality. Here we investigate the minimal values $W_{\rm min}^b$ of the Wigner function $W_b(\xi_{b1}, \xi_{b_2})$ of the mechanical state $\hat \rho_b^{\rm ss}$, with the definition
\begin{align}
W_b(\xi_{b1}, \xi_{b_2})=&\frac{1}{\pi^4 } \int_{-\infty }^{\infty } \mathrm{d}^2 \eta_{b_1}  \mathrm{d}^2\eta_{b_2} \chi \left ( \eta_{b_1} , \eta_{b_2}  \right )\nonumber\\
&\times e^{ ( \xi_{b_1} \eta_{b_1} ^{\ast } -\xi_{b_1} ^{\ast } \eta_{b_1} + \xi_{b_2}  \eta_{b_2}  ^{\ast } -\xi_{b_2}  ^{\ast } \eta_{b_2} )} ,
\end{align}
where $\chi \left ( \eta_{b_1} , \eta_{b_2}  \right )\equiv Tr\big[\hat \rho_b^{\rm ss}\hat D(\xi_{b_1})\hat D(\xi_{b_2})\big]$ is the characteristic function of the mechanical state $\hat \rho_b^{\rm ss}$, with the displacement operators $D(\xi_{b_1})$ and $\hat D(\xi_{b_2})$.
In addition to the Wigner negativity, we also investigate the entanglement in the mechanical state, which can be witnessed the negativity \cite{eng}
\begin{align}
\mathcal N(\hat \rho_b^{\rm ss})= Tr\big(\sqrt{\hat \rho_{b, PT}^{\rm ss \dagger} \hat \rho_{b, PT}^{\rm ss }}-1\big)/2,
\end{align}
where $\hat \rho_{b, PT}^{\rm ss }$ denotes the state after the partial transpose on the mechanical state $\hat \rho_b^{\rm ss}$.
In Fig.\ref{fig_sprop} (b) and (c), we plot the minimal values $W_{\rm min}^b$ and the partial transpose negativity $\mathcal N$. It is shown that similar to the non-Gaussianity, the Wigner negativity and entanglement also increase with the parameter $\zeta$ for both pure and realistic mechanical PCSs, in spite of the deceasing of the fidelity with $\zeta$ for the state $\hat \rho_b^{\rm ss}$ in Fig.\ref{fig_sprop} (c). The Wigner negativity and the entanglement exhibit the similar dependences on $\zeta$, and both are also diminished by the mechanical damping and counter-rotating terms, compared to those in Fig.\ref{fig_sprop} (c).

We also investigate the properties of non-Gaussian EPR steering correlations between the two mechanical oscillators. We consider the generalized Reid criterion \cite{er1,er2,er3} to witness the non-Gaussian quantum steering, given by
\begin{align}
E_r^{(m,n)}=\frac{2\sqrt{ \big\langle V( \hat X_{1}^{(m)}|  x_{2}^{(n)})  \big\rangle_e\big\langle V( \hat Y_{1}^{(m)}|  y_{2}^{(n)})  \big\rangle_e } }{\left | \left \langle \left [  \hat X_{1}^{(m)},\hat Y_{1}^{(m)}\right ] \right \rangle_{\hat \rho_{b_1}}   \right |},
\end{align}
where the $m$-order quadratures $\hat X_{1}^{(m)}$ and $\hat Y_{1}^{(m)}$ of the first mechanical oscillator are defined as $\hat X_{1}^{(m)}=(\hat{b}_1^m +\hat{b}_1^{\dagger m})/2$ and $\hat Y_{1}^{(m)}=-i(\hat{b}_1^m -\hat{b}_1^{\dagger m})/2$. $ V( \hat X_{1}^{(m)}| x_{2}^{(n)})$ is the conditional variance of $\hat X_{1}^{(m)}$ on the outcome $ x_{2}^{(n)}$ of the measurement of $n$-order quadrature $\hat X_{2}^{(n)}$ on the second mechanical oscillator $\hat b_2$, and $\big\langle V( \hat X_{1}^{(m)}| x_{2}^{(n)})  \big\rangle_e\equiv\int \mathrm{d}x_{2}^{(n)} P ( x_{2}^{(n)}  ) V( \hat X_{1}^{(m)}| x_{2}^{(n)})$ represents the ensemble average of $ V( \hat X_{1}^{(m)}| x_{2}^{(n)}) $ over all the possible outcomes $ x_{2}^{(n)}$ ($\big\langle V( \hat Y_{1}^{(m)}|  y_{2}^{(n)})  \big\rangle$ is defined similarly). The operator $\hat \rho_{b_1}$ is the reduced density operator of the first mechanical oscillator, i.e., $\hat \rho_{b_1}=Tr_{b_2}[\hat \rho_{b}]$.
The condition $E_r^{(m,n)}<1$ serves as the evidence of the steering from the second mechanical oscillator to the first one.

In Fig.\ref{fig_sprop}~(d) and (e), $E_r^{(m,n)}$ is plotted, respectively, for the pure and approximate mechanical PCSs. For the pure state, we find that only when $m=n=1,2$, the steering can be detected. When $m=n=1$, the steering increases at first and then decreases with parameter $\zeta$, different from the dependence of the mechanical entanglement. The higher-order steering for $m=n=2$ is observed when $\zeta>0.7$. However, for the approximate mechanical state, with the Reid criterion, the evidence of the steering is absent in both cases.

To better reveal the quantum steering of the mechanical PCSs, we consider the metrological characterization of the non-Gaussian steering with quantum Fisher information \cite{fi5,fi4}. The metrological protocol for detecting the quantum steering is as follows:
for the two mechanical modes, we assume that the observer Alice perform homodyne detection on the mechanical mode $\hat b_2$, and then the other observer Bob perform a phase shift via the displacement operation on the mechanical mode $\hat b_1$, i.e., $ \hat\rho_{b_1}^{\xi|\rm con} =e^{-i\hat X_{1} \xi } \hat \rho_{b_1}^{\rm con} e^{i\hat X_{1} \xi }$, to encode the parameter $\xi$ to be estimated, where $\hat \rho_{b_1}^{\rm con}$ is the conditional state of the mechanical mode $\hat b_1$, and estimate the parameter $\xi$ via homodyne detection.
The quantum Fisher information (QFI) characterizes the optimal precision that Bob can achieve in estimating $\xi$ by directly measuring the state $\hat \rho _{b}^{\xi}$.
The classical Fisher information (CFI), which corresponds to the specific homodyne detection, is the lower bound of the QFI \cite{fi1,fi2,fi3} and defined as
$F_{\xi } \left [ P \right ] = \int\mathrm{d}x_1 P\left ( x_1|\xi  \right ) \left ( \frac{\partial \mathcal{L}\left ( x_1|\xi  \right ) }{\partial \xi}   \right )^2,$
where $P\left ( x_1|\xi  \right )
$ denotes the probability distribution of the measurement outcome $x_1$ for the observable $\hat X_1$ and $
\mathcal{L}\left ( x_1|\xi  \right )= \log\left [ P\left ( x_1|\xi  \right ) \right ]
$.
In our two-mode entangled state, Alice can assist Bob to enhance his estimation precision of \(\xi\) by performing the homodyne detection with the observable $\hat X_{2\theta}=\hat b_2 e^{-i\theta}+\hat b_2^{\dagger} e^{i\theta}$ and the outcome \( x_{\theta2} \).
The resulting CFI, optimized over $ x_{\theta2}$, is expressed as
$F^{hom}_{b_1|b_2}  =  \displaystyle \max_{\theta \in  [ 0,2\pi )}   \int \mathrm{d}x_{\theta2} P ( x_{\theta2}   ) F_{\xi }  [ P^{x_{\theta2}}_{b_1} (x_1|\xi)  ] $, where  $P ( x_{\theta2}   )$ is the probability distribution of $ x_{\theta2}$ and the conditional probability $
P^{x_{\theta2} }_{b_1} \left ( x_1|\xi \right ) = P^{x_{\theta2}}_{b_1} \left ( x_1-\xi \right )$, dependent on the detection results $x_{\theta2} $.
The conditional variance
$V^{\rm hom}_{b_1|b_2}  =\displaystyle \min_{\theta \in  [ 0,2\pi )} \langle V (\hat X_1| x_{\theta2})\rangle= \displaystyle \min_{\theta \in  [ 0,2\pi )} \int \mathrm{d}x_{\theta2} P ( x_{\theta2}   ) V (\hat X_1| x_{\theta2}  )$. When the steering is present from the second mechanical oscillator to the first one, the steering criterion in terms of the Fisher information is expressed as \cite{fi4}
\begin{align}
E_f= \max\left[0,F^{hom}_{b_1|b_2} - 4V^{hom}_{b_1|b_2}\right].
\end{align}
When the condition $E_f>0$ holds, the conditional estimation precision exceeds the bound imposed by the uncertainty principle, thus proving the presence of steering.

The results are presented in Fig.\ref{fig_sprop}~(d) and (e). Fig.\ref{fig_sprop}~(d) shows that through the metrological characterization, the steering of the pure PCS increases monotonically with $\zeta$, different from the dependence behavior of the steering revealed with the Reid criterion, but the same as the dependence of the entanglement and Wigner negativity. Moreover, as illustrated in Fig.\ref{fig_sprop}~(e), for the approximate mechanical PCS, although the steering becomes weaker, it is present and also has a tendency to increase with the parameter $\zeta$, consistent to the entanglement and Wigner negativity. The Reid criterion fails to witness the steering of the approximate mechanical PCS. Therefore, this shows clearly that the fisher-information-based criterion is more efficient for detecting the steering of the non-Gaussian mechanical state.
It is noteworthy that for the approximate mechanical PCS, the steering  is absent for $\zeta <0.8$, as shown in Fig.\ref{fig_sprop}~(e), which is due to the decreasing of the fidelity.



\section{Remote generation of motional cat states with the non-Gaussian mechanical entangled states}
\begin{figure*}[t]
\centerline{\scalebox{0.38}{\includegraphics{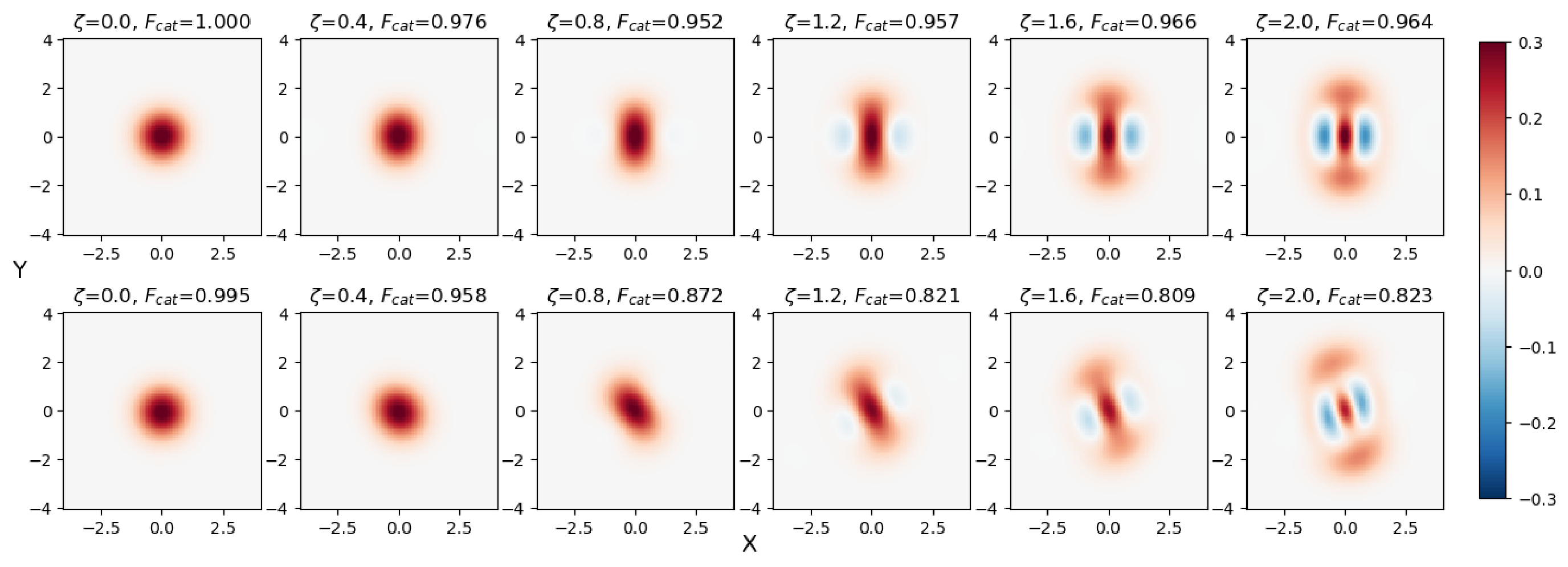}}}
\caption{The Wigner functions of the Schr\"{o}dinger cat states generated with the pure (first row) and approximate (second row) mechanical PCSs in Fig.2 (a), for different values of the parameter $\zeta$ and the corresponding fidelity $F_{\rm cat}$ of the cat states.}
\label{fig_cat}
\end{figure*}
In this section, we study the remote preparation of mechanical Schr\"{o}dinger cat states via the achieved non-Gaussian mechanical entanglement. As the above the metrological protocol, when the mechanical mode $\hat b_2$ is homodyne detected, with the specific result $x_2=0$, the conditional state of the mechanical mode \( \hat b_1 \) is given by
\begin{align}
\left |  \zeta  \right \rangle_{b_1}^{x_2=0}
\propto  \sum_{n=0}^{\infty }
 \frac{\zeta ^{n} }{n!} \langle   0 |  n   \rangle_2| n \rangle_1.
\end{align}
With the expression  $\langle   0 |  n   \rangle_2=\frac{H_{n}(0)  }{\sqrt{2^{n}n! } }$, where $H_{n}(0)$ is the Hermite polynomial of order $n$, we have
\begin{align}
\left |  \zeta  \right \rangle_{b_1}^{x_2=0}
\propto \sum_{n=0}^{\infty }\frac{\zeta ^{n} }{n!}\frac{H_{n}(0) }{\sqrt{2^{n}n! } } | n \rangle_1.
\end{align}
Considering $H_{2k+1}(0)= 0$ and $H_{2k}(0) =(-2)^k(2k-1)!!$, the conditional state becomes
\begin{align}
\left |  \zeta  \right \rangle_{b_1}^{x_2=0}
\propto \sum_{k=0}^{\infty }\frac{(i\zeta) ^{2k} }{\sqrt{(2k)!}}\frac{(\frac{k}{2}) ^{k} }{k!} | 2k \rangle_1.
\end{align}
Under the condition that $\zeta$ is relatively large, $\frac{(\frac{k}{2}) ^{k} }{k!} $ varies much more smoothly with respect to $k$, compared to $\frac{(i\zeta) ^{2k} }{\sqrt{(2k)!}}$. As a result, the conditional state of the mechanical mode $\hat b_1$ mode can be further approximated as
\begin{align}
\left |  \zeta  \right \rangle_{b_1}^{x_2=0}
\propto \sum_{k=0}^{\infty }\frac{(i\zeta) ^{2k} }{\sqrt{(2k)!}} | 2k \rangle_1
\propto \left | -i\sqrt{\zeta }   \right \rangle +\left | i\sqrt{\zeta }   \right \rangle .
\end{align}
This corresponds to an even cat state with amplitude $\sqrt{\zeta }$. 
Thus, as we see, through the homodyne detection on the mechanical mode $\hat b_2$, the odd number states are filtered out from the conditional state in Fock space when the detection outcome $x_2=0$, and the even cat state can thus be generated. This embodies the essence of the non-Gaussian quantum steerable correlations in the PCS, since for a two-mode Gaussian state (e.g., a two-mode squeezed vacuum) the homodyne detection on one mode merely results in conditional Gaussian state of the other mode. Therefore, in the Gaussian setting, to remote prepare negative Wigner states via homodyne detection, one has to at first perform non-Gaussian operation, e.g., photon addition or subtraction, to convert Gaussian states to non-Gaussian ones conditionally\cite{hesu,mp16}. But for the present scheme, the steady-state non-Gaussian mechanical state is prepared unconditionally with the cavity dissipation and nonlinear optomechanical coupling. We further note that when the initial phonon-number difference between the two mechanical oscillators is accurately controlled to be odd, the homodyne detection can also lead to odd cat states.


To achieve homodyne detect on the second mechanical oscillator, we consider a state transfer scheme in which the mechanical oscillator is weakly coupled to the other cavity, denote by $\hat a_t$, as shown in Fig.1~(b). When the two mechanical oscillators cavity reaches the steady state, the cavity $\hat a_t$ is driven by a red-detuned red pulsed laser to realize the state transfer from the second mechanical oscillator to the cavity output field\cite{syst1}. The Hamiltonian in the rotating frame of the drive frequency \( \omega_{p_2} \) is given
\begin{align}
\hat H_t=&\Delta_{t}\hat{a}_t^{\dagger}\hat{a}_t+\omega_{b_2}\hat{b}_2^{\dagger}\hat{b}_2+g_t \hat{a}_t^{\dagger}\hat{a}_t(\hat{b}_2+\hat{b}_2^{\dagger})\nonumber\\
&+\left(\varepsilon_{p_2}\hat{a}_t^{\dagger}+\varepsilon_{p_2}^{*}\hat{a}_t \right),
\label{Ht}
\end{align}
where $\Delta_{t}=\omega_{t}-\omega_{p_2}$, $\omega_{t}$ is the cavity resonant frequency, $g_t$ is the single-photon optomechanical coupling rate, and
$\varepsilon_{p_2}$ is the drive amplitude. By displacing the cavity by its amplitude $\alpha_t = \frac{\varepsilon_{p_2}}{i\gamma_{t} - \Delta_t}$, where $\gamma_{t}$ is the dissipation rate of the cavity field $\hat a_t$, the Hamiltonian (\ref{Ht}) can be approximately linearized into
\begin{align}
\hat H_t&=i\widetilde{g}\hat{b}_2^{\dagger }\hat{a}_t-i\widetilde{g}^{*}\hat{b}_2\hat{a}_t^{\dagger},
\end{align}
with the coupling $\widetilde{g}= -ig_t\alpha_t^{*}$, under conditions $\Delta_t = \omega_{b_2}$ [as depicted in Fig.1 (d)] and $\omega_{b_2}\gg \{\widetilde{g},\gamma_t\}$.

The time evolution of the operators $\hat a_t$ and $\hat b_2$ can be derived as
$\frac{d}{dt}\hat {a}_{t}
=-\widetilde{g}\hat{b}_2-\gamma_t \hat a_{t}-\sqrt{2\gamma_t}\hat a_{\rm in}$
and
$\frac{d}{dt}\hat {b}_{2}=\widetilde{g}\hat a_{t}$, where $\hat a_{\rm in}$ denotes the vacuum fluctuations entering the cavity and the mechanical damping can be neglected when the pulse duration $\tau\ll (\gamma_{b_2}\bar {n}_{b_2})^{-1}$.
Under the condition of $\widetilde{g}\ll\gamma_t$, we can adiabatically eliminate the cavity field and
\begin{align}
\hat B_{\rm out} &= e^{-G\tau}\hat B_{\rm in}-\sqrt{1-e^{-2G\tau}}\hat A_{\rm in},
\end{align}
where
$G = \frac{\widetilde{g}^2}{\kappa},\quad \hat B_{in}=\hat{b}_2(0),\quad \hat B_{\rm out}= \hat{b}_2(\tau),\quad \nonumber\\
\hat A_{in} = \sqrt{\frac{2G}{e^{2Gt}-1}}\int_{0}^{\tau}\hat a_{in}(t)e^{Gt}dt$.
Considering the input-output relationship $\hat a_{\rm out}=\hat a_{\rm in}+\sqrt{2\kappa}\hat{a}_t$ can be obtained
\begin{align}
\hat A_{\rm out} &= e^{-G\tau}\hat A_{\rm in}-\sqrt{1-e^{-2G\tau}}\hat B_{\rm in}.
\end{align}
When adjusting the pulse duration $\tau$ to satisfy $e^{-G\tau}\rightarrow 0$, we have
$\hat A_{out}\simeq-\hat B_{in}$, which means that the state $\hat \rho^{\rm ss}_b$ of the two mechanical oscillators is mapped onto the first mechanical oscillator and the probe cavity output in the state $\hat \rho_t\simeq\hat \rho^{\rm ss}_b$. This allows us remotely manipulate the mechanical states via local measurement with the help of the nonlocal quantum optomechanical correlations exhibited in $\hat \rho_t$.

We now consider homodyne detection on the output $\hat A_{\rm out}$, which gives
the conditional state of the first mechanical oscillator as
\begin{align}
\hat \rho_{b_1}^{\rm con}\left ( x_{a_t} \right ) & = \frac{ Tr_{a_t}\left [  \left ( \hat M_{a_t}\otimes \hat I_{b_1} \right )\hat \rho_{t}  \left ( \hat I_{b_1}\otimes \hat M_{a_t} \right )\right ] }{Tr_{b_1} \left \{ Tr_{a_t}\left [  \left ( \hat M_{a_t}\otimes \hat I_{b_1} \right )\rho_{t}  \left ( \hat I_{b_1}\otimes \hat M_{a_t} \right )\right ] \right \}},
\end{align}
where $\hat M_{a_t} = \left | x_{a_t} \right \rangle   \left \langle x_{a_t} \right |$ is the projection operator of the eigenstate $|x_{a_t}\rangle$ of the quadrature operator $\hat X_{a_t}  = \hat A_{\rm out} + \hat A_{\rm out}^{\dagger}$.

In Fig.\ref {fig_cat}, the Wigner function of the state $\hat \rho_{b_1}^{\rm con}(x_{a_t}=0)$ is displayed for different values of the parameter $\zeta$ and fidelity $F_{\rm cat}$ when the detection result $x_{a_t} =0$. We consider the parameters $\gamma_{t}\approx0.1\omega_{t}$, $g_{t}\approx0.1\gamma_{t}$, $\gamma_{t}\approx\varepsilon_{p_2}$ , and the duration $\tau\approx 2\times10^{2}\omega_{b_2}^{-1}\ll800\gamma_a^{-1}\approx5\times10^{5}\omega_{b_2}^{-1}$.
The first and second rows correspond to the pure and approximate mechanical PCSs in Fig.2~(a), respectively.
It is evident that as \( \zeta \) increases, the conditional Wigner functions of exhibit increasingly pronounced characteristics of even cat states. With the approximate mechanical PCS, the fidelity of the motional cat states decreases as $\zeta$ increases, due to the fact that the fidelity of the mechanical state $\hat \rho^{\rm ss}_b$, with respect to the pure PCS, also decreases with $\zeta$. For the amplitude $\zeta=2.0$, the optimal fidelity of the cat state is 0.82, showing the present scheme is efficient for realizing the remote generation of the mechanical cat states.

\section{Effects of thermal fluctuations}
\begin{figure}
\centerline{\scalebox{0.082}{~~\includegraphics{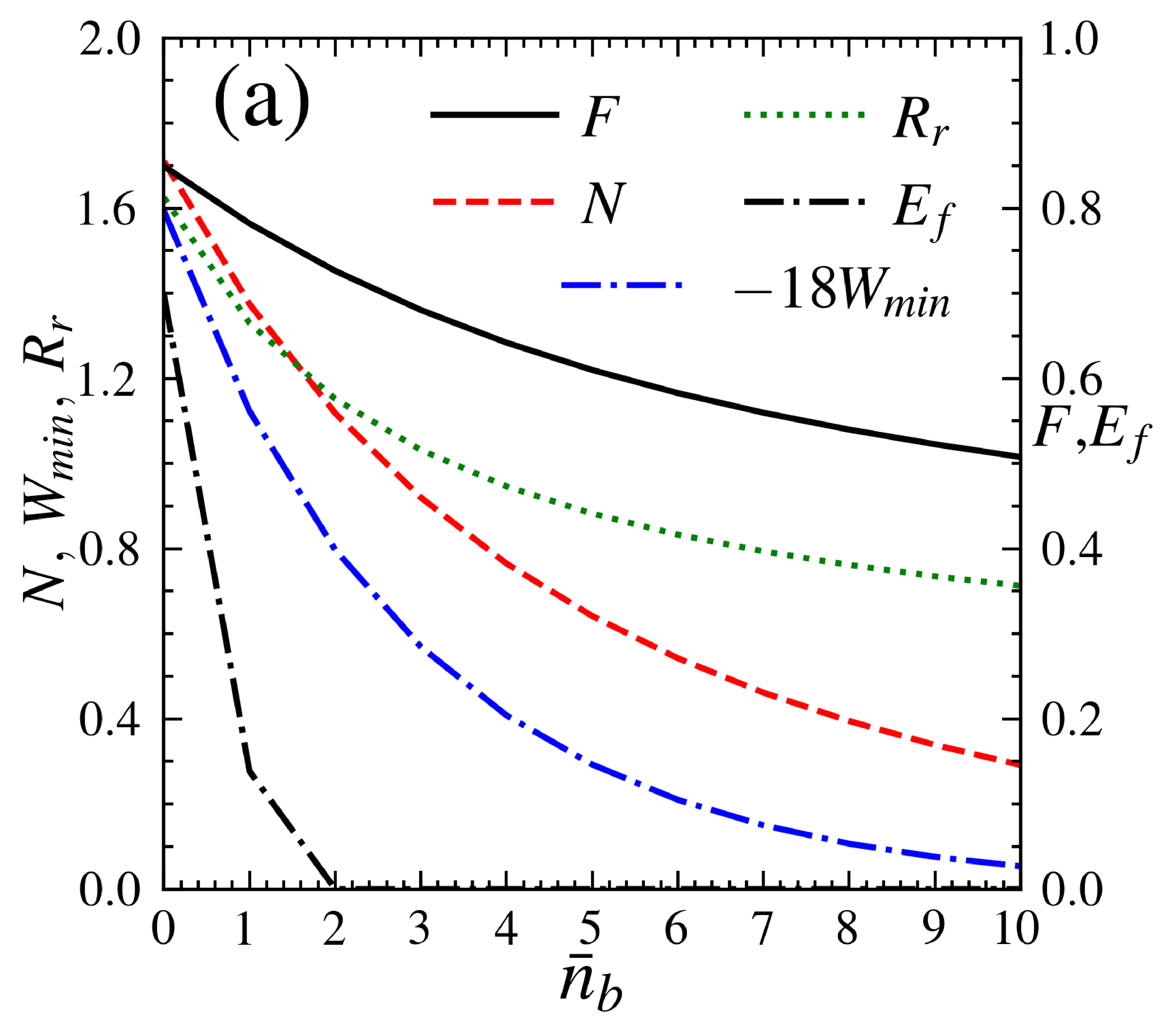}\hspace{-0.8cm}\includegraphics{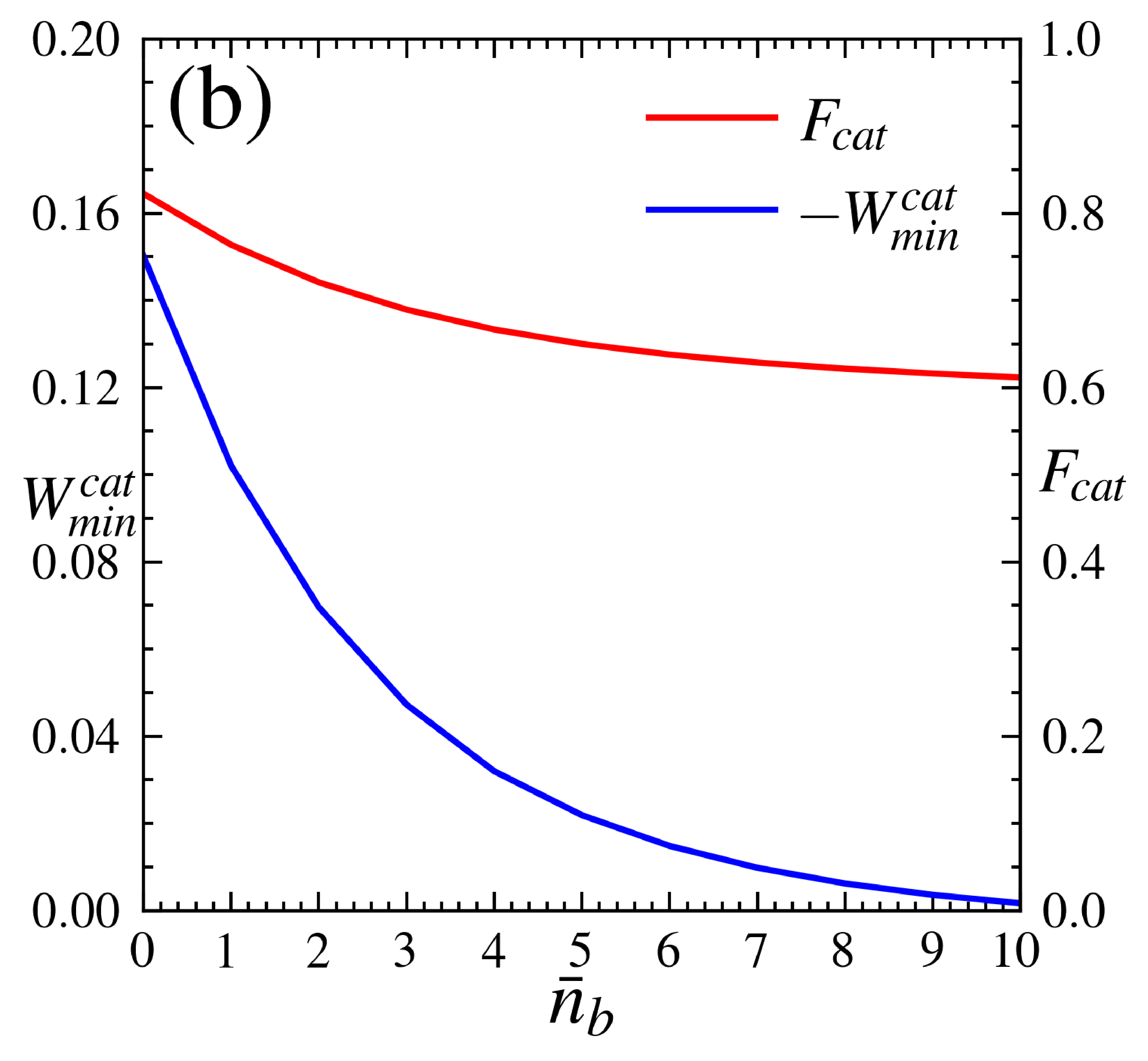}}}
\vspace{-5pt}
\caption{The dependence of the fidelity $F$, non-Gaussianity $R_r$, Wigner negativity $W_{\rm min}$, mechanical entanglement $\mathcal N$ and steering $E_f$ for the mechanical entangled states on the thermal mean phonon number $\bar{n}_{b}\equiv\bar{n}_{b_j}$.
(b) The fidelity $F_{\rm cat}$ and Wigner negativity $W_{\rm min}^{\rm cat}$ of the mechanical cat state as the functions of \( \bar{n}_b \).}
\label{fig_nb}
\end{figure}

Finally, we estimate the effects of the thermal fluctuations on the mechanical states. We still numerically solving the full master equation (\ref{ms_o}) by utilizing the parameters given in Sec.II but with $\bar {n}_{b_j}=\bar {n}_{b}\neq0$.
Our numerical results are presented in Fig. \ref{fig_nb} for the parameter $\zeta =2$.
As shown in Fig.\ref{fig_nb}~(a), as the thermal excitation number $ \bar{n}_{b}$ increases, the non-gaussianity, entanglement, and Wigner negativity of the mechanical PCS decreases. The Wigner negativity is more sensitive to the thermal fluctuations than the other two properties: both entanglement and non-Gaussianity can persist up to $\bar{n}_{b}>10$, whereas the Wigner negativity vanishes nearly at $\bar{n}_{b}=10$. Further, the quantum steering is much more susceptible to the thermal noise, with the steerability being completely suppressed around
$\bar{n}_{b}=2$. Fig.\ref{fig_nb}~(b) reveals that the conditional mechanical state $\hat \rho_{b_1}^{\rm con}(x_{a_t}=0)$ maintains the Wigner negativity $W_{\rm min}^{\rm cat}$ up to $\bar{n}_{b}=10$, at which the fidelity $F_{\rm cat}$ of the cat state degrades to 0.61 from the maximum 0.82. For instance, when considering the mechanical frequencies $\omega_{b_j}/2\pi\sim 5 \mathrm {GHz} $ as in Ref. \cite{exp2}, the Wigner negativities can be achieved up to the temperature $T\sim2.5 \mathrm {K}$, with $\bar{n}_{b_j}=(e^{\hbar \omega_{b_{j}} / k_B T}- 1)^{-1}$, where $k_B$ is the Boltzmann constant.

\vspace{-10pt}
\section{Conclusion}
\vspace{-10pt}
In conclusion, we propose a driven-dissipative scheme for generating steady non-Gaussian mechanical entangled states of two mechanical oscillators and consider the remote preparation of Schr\"{o}dinger cat states of one mechanical oscillator via the mechanical non-Gaussian entanglement and homodyne detection. We consider a cavity optomechanical system consisting two mechanical oscillators of different frequencies inside a bichromatically-driven cavity. We show that an effective nondegenerate parametric downconversion involving the mechanical oscillators and cavity field can be engineered under appropriate conditions. We analytically and numerically reveal that the cavity dissipation drives the two mechanical oscillators into a steady PCS.  The properties of the mechanical non-Gaussian state, including no-Gaussianity, the Wigner negativity, entanglement, and quantum steering are studied in detail. We demonstrate that homodyne detection on one mechanical oscillator can project the other oscillator into a Schr\"{o}dinger cat state with the help of the non-Gaussian mechanical entanglement. The detection can be realized, as we show, by transferring the mechanical state to the output field of an auxiliary probe cavity coupled to the target oscillator and subsequently homodyning on the output field. We also discuss the robustness of the achieved mechanical entangled states against thermal fluctuations. Our findings establish a feasible approach for the dissipative and remote preparation of mechanical nonclassical states and may be applicable to fundamental test of quantum physics and quantum tasks, such as quantum-enhanced metrology with the mechanical PCS.


\vspace{-10pt}
\section*{Acknowledgment}
\vspace{-10pt}
This work is supported by the National Natural Science Foundation of China (No. 12174140).

\begin{widetext}
\section*{APPENDIX: The detailed derivation of Eqs.(\ref{h_a}), (\ref{ms_b}) and (\ref{pcs})}

As shown in Fig.\ref{sys1}~(a), we consider a cavity optomechanical system in which the cavity field dispersively interacts with two mechanical oscillators and is simultaneously driven by both a strong and a weak laser. The interaction of the system can be described by the Hamiltonian
\begin{align}
\hat H_1=\omega_{c}\hat{a}^{\dagger}\hat{a}+\sum_{j=1}^{2}\omega_{b_j}\hat{b}_j^{\dagger}\hat{b}_j+g_j\hat{a}^{\dagger}\hat{a}\big(\hat{b}_j+\hat{b}_j^{\dagger}\big)+\left(\varepsilon_{p}\hat{a}^{\dagger}e^{-i\omega_{p}t}+\varepsilon_{d}\hat{a}^{\dagger}e^{-i\omega_{d}t}+h.c.\right),
\end{align}
where the annihilation operator $\hat{a}~(\hat{b}_j)$ denotes the cavity field (mechanical oscillator) with resonant frequency $\omega_{c}$ ($\omega_{b_j}$), $g_j$ are the single-photon optomechanical coupling rates, and $\omega_{p}$~($\varepsilon_p$) and $\omega_{d}$ ($\varepsilon_{d}$) are frequencies~(amplitudes) of the drive fields with the relations
\begin{align}
\omega_d=\omega_p+\omega_{b_1}+\omega_{b_2}, ~~\omega_d\approx \omega_c,~~\varepsilon_p\gg\varepsilon_d,
\label{reltA}
\end{align}
as depicted in Fig.\ref{sys1}~(b).

Including the cavity dissipation and mechanical damping, the density operator $\hat \rho_{ab}$ of the system is satisfied by the master equation
\begin{align}
\frac{d}{dt}{\hat \rho_{ab}} = -i[\hat H_1,\hat \rho_{ab}]+\gamma_{a}\mathcal L[\hat{a}]\hat \rho_{ab}+\sum_j\gamma_{b_j}(\bar{n}_{b_j}+1)\mathcal L[\hat{b}_j]\hat \rho_{ab}
+\gamma_{b_j}\bar{n}_{b_j}\mathcal L[\hat{b}_j^\dagger]\hat \rho_{ab},
\label{ms_oA}
\end{align}
where $\gamma_a$ is the dissipation rate of the cavity field in a vaccum bath and $\gamma_{b_j}$ are the damping rates of the mechanical oscillators in their thermal environments with the mean thermal photon numbers $\bar{n}_{b_j}$.
We assume the condition $\gamma_{a} \gg \gamma_{b_j}$ holds, which allows us to temporarily neglect the mechanical dissipation terms in the subsequent derivation. By adopting the drive frequency $\omega_{p}$ as the rotating frame, equivalent to applying the unitary transformation $e^{i\omega_{p}\hat{a}^{\dagger}\hat{a}t}$ to the master equation (\ref{ms_oA}):
\begin{align}
e^{i\omega_{p}\hat{a}^{\dagger}\hat{a}t}\frac{d}{dt}\hat \rho_{ab} e^{-i\omega_{p}\hat{a}^{\dagger}\hat{a}t}
=& e^{i\omega_{p}\hat{a}^{\dagger}\hat{a}t}\left\{-i\left[\hat H_1,\hat \rho_{ab}\right]+\gamma_{a}\mathcal L[\hat{a}]\hat \rho_{ab}\right\}e^{-i\omega_{p}\hat{a}^{\dagger}\hat{a}t}\nonumber\\
e^{i\omega_{p}\hat{a}^{\dagger}\hat{a}t}(\frac{d}{dt}\hat \rho_{ab}+i\omega_{p}\hat{a}^{\dagger}\hat{a}\hat \rho_{ab} -i\omega_{p}\hat \rho_{ab} \hat{a}^{\dagger}\hat{a})e^{-i\omega_{p}\hat{a}^{\dagger}\hat{a}t}
=& e^{i\omega_{p}\hat{a}^{\dagger}\hat{a}t}\Big\{-i\left[\hat H_1-\omega_{p}\hat{a}^{\dagger}\hat{a},\hat \rho_{ab} \right]+\gamma_{a}
\mathcal L[\hat{a}]\hat \rho_{ab}\Big\}e^{-i\omega_{p}\hat{a}^{\dagger}\hat{a}t}\nonumber\\
\frac{d}{dt}(e^{i\omega_{p}\hat{a}^{\dagger}\hat{a}t}\hat \rho_{ab} e^{-i\omega_{p}\hat{a}^{\dagger}\hat{a}t})
=& -i\left[e^{i\omega_{p}a^{\dagger}\hat{a}t}(\hat H_1-\omega_{p}\hat{a}^{\dagger}\hat{a})e^{-i\omega_{p}\hat{a}^{\dagger}\hat{a}t}, \hat \rho_{ab}^{u_1} \right] + \gamma_{a}\mathcal L[\hat{a}]\hat \rho_{ab}^{u_1}\nonumber\\
\frac{d}{dt} \hat \rho_{ab}^{u_1}
=&-i\left[ \hat{\widetilde{H}_1}, \hat \rho_{ab}^{u_1} \right]+ \gamma_{a}\mathcal L[\hat{a}]\hat \rho_{ab}^{u_1}
\label{ms_w}
\end{align}

To distinguish the density matrix and Hamiltonian after the unitary transformation, we set $\hat \rho_{ab}^{u_1}=e^{i\omega_{p}\hat{a}^{\dagger}\hat{a}t}\hat \rho_{ab} e^{-i\omega_{p}\hat{a}^{\dagger}\hat{a}t}$ and $\hat{\widetilde{H}_1}=e^{i\omega_{p}\hat{a}^{\dagger}\hat{a}t}(\hat H_1-\omega_{p}\hat{a}^{\dagger}\hat{a}) e^{-i\omega_{p}\hat{a}^{\dagger}\hat{a}t}$. And $\hat \rho_{ab}^{u_1}$ and $\hat \rho_{ab}$ represent different representations of the same state. It can be observed that the dissipation operator $\hat{a}$ of the dissipation term $\gamma_a\mathcal L[\hat{a}]\rho_{ab}^{u_1}$ remains unchanged, still corresponding to single-photon dissipation. By substituting $\hat e^{i\omega_{p}\hat{a}^{\dagger}\hat{a}t}\hat a e^{-i\omega_{p}\hat{a}^{\dagger}\hat{a}t}=a e^{-i\omega_{p}t}$, the Hamiltonian $\hat{\widetilde{H}_1}$ is obtained as :
\begin{align}
\hat{\widetilde{H}_1}=\Delta \hat{a}^{\dagger}\hat{a}+\sum_{j=1}^{2}\omega_{b_j}\hat{b}_j^{\dagger}\hat{b}_j+g_j\hat{a}^{\dagger}\hat{a}\big(\hat{b}_j+\hat{b}_j^{\dagger}\big)+(\varepsilon_{p}a^{\dagger}+\varepsilon_{d}a^{\dagger}e^{i\Delta pt}+h.c.)
\end{align}
where the detuning $\Delta = \omega_{c} - \omega_{p}$ and $\Delta_{p}= w_{p}-w_{d}$.
By displacing the cavity by its steady state amplitude $\alpha = \frac{\varepsilon_{p}}{i\gamma_a - \Delta}$, which considers only the drive field with frequency $\omega_p$,  equivalent to applying the unitary $e^{\alpha^{*}\hat{a}-\alpha \hat{a}^{\dagger}}$ to the master equation (\ref{ms_w}). Under this unitary transformation, the operators are transformed as follows:
\begin{align}
e^{\alpha ^{*}\hat{a}-\alpha \hat{a}^{\dagger }} \hat{a}^{\dagger } e^{\alpha \hat{a}^{\dagger }-\alpha ^{*}\hat{a}}  &= \hat{a}^{\dagger } + \left[ \alpha \hat{a}^{\dagger }-\alpha ^{*}\hat{a} , \hat{a}^{\dagger}\right]+\left[ \alpha \hat{a}^{\dagger }-\alpha ^{*}\hat{a} , \left[ \alpha \hat{a}^{\dagger }-\alpha ^{*}\hat{a} , \hat{a}^{\dagger}\right] \right]+...  = \hat{a}^{\dagger } + \alpha ^{*}\nonumber\\
e^{\alpha ^{*}\hat{a}-\alpha \hat{a}^{\dagger }} \hat{a}^{\dagger }\hat{a} e^{\alpha \hat{a}^{\dagger }-\alpha ^{*}a} & = a^{\dagger }a + \alpha^{*}\hat{a} + \alpha a^{\dagger} + \left | \alpha \right | ^{2}
\end{align}
Substituting into the Hamiltonian $\hat{\widetilde{H}_1}$, we obtain:
\begin{align}
e^{\alpha^{*}\hat a-\alpha \hat a^{\dagger}}\hat{\widetilde{H}_1}e^{\alpha \hat a^{\dagger }-\alpha ^{*}\hat a}=&\Delta(\hat a^{\dagger }\hat a+\alpha^{*}\hat a+\alpha \hat a^{\dagger}+|\alpha|^{2})+\omega_{ b_1}\hat b_{1}^{\dagger}\hat b_{1}+\omega_{b_2}\hat b_{2}^{\dagger}\hat b_{2}+(\varepsilon_{p}a^{\dagger}+\varepsilon_{d}a^{\dagger}e^{i\Delta pt}+h.c.)\nonumber\\
&+(\hat a^{\dagger }\hat a+\alpha^{*}\hat a+\alpha \hat a^{\dagger }+|\alpha|^{2})(g_{1}\hat b_{1}+g_{1}\hat b_{1}^{\dagger}+g_{2}\hat b_{2}+g_{2}\hat b_{2}^{\dagger})
\end{align}
The dissipation term after the unitary transformation becomes:
\begin{align}
e^{\alpha^{*}\hat a-\alpha \hat a^{\dagger}} \mathcal L[\hat a]\hat \rho_{ab}^{u_1} e^{\alpha \hat a^{\dagger }-\alpha ^{*}\hat a}
=& 2(\hat a+\alpha) \hat \rho_{ab}^{u_2} (\hat a^{\dagger }+\alpha^{*})-(\hat a^{\dagger }+\alpha^{*})(\hat a+\alpha)\hat \rho_{ab}^{u_2}-\hat \rho_{ab}^{u_2}(\hat a^{\dagger }+\alpha^{*})(\hat a+\alpha)\nonumber\\
=&2\hat a\hat \rho_{ab}^{u_2}\hat a^{\dagger }+2\hat a\hat \rho_{ab}^{u_2}\alpha ^{*}+2\alpha \hat \rho_{ab}^{u_2}\hat a^{\dagger } + 2|\alpha|^{2}\hat \rho_{ab}^{u_2}\nonumber\\
&- (\hat a^{\dagger }\hat a+\alpha^{*}\hat a+\alpha \hat a^{\dagger }+|\alpha|^{2})\hat \rho_{ab}^{u_2}-\hat \rho_{ab}^{u_2}(\hat a^{\dagger }\hat a+\alpha^{*}\hat a+\alpha \hat a^{\dagger }+|\alpha|^{2})\nonumber\\
=&2\hat a\hat \rho_{ab}^{u_2}\hat a^{\dagger }-\hat a^{\dagger }\hat a\hat \rho_{ab}^{u_2}-\hat \rho_{ab}^{u_2}\hat a^{\dagger }\hat a+\hat a\hat \rho_{ab}^{u_2}\alpha ^{*}+\alpha \hat \rho_{ab}^{u_2}\hat a^{\dagger }-\alpha \hat a^{\dagger }\hat \rho_{ab}^{u_2}-\alpha^{*}\hat \rho_{ab}^{u_2}\hat a\nonumber\\
=&2\hat a\hat \rho_{ab}^{u_2}\hat a^{\dagger }-\hat a^{\dagger }\hat a\hat \rho_{ab}^{u_2}-\hat \rho_{ab}^{u_2}\hat a^{\dagger }\hat a-i[-i(\alpha \hat a^{\dagger }-\alpha^{*}\hat a),\hat \rho_{ab}^{u_2}]
\end{align}
Where $\hat \rho_{ab}^{u_2}=e^{\alpha^{*}\hat a-\alpha \hat a^{\dagger}}\hat \rho_{ab}^{u_1} e^{\alpha \hat a^{\dagger }-\alpha ^{*}\hat a}$. Since $\alpha = \frac{\varepsilon_{p}}{i\gamma_a - \Delta}$, the relation $\Delta(\alpha^{*} \hat a+\alpha \hat a^{\dagger})
+\varepsilon_{p}\hat a^{\dagger}+\varepsilon_{p}^{*}\hat a-i\gamma_a(\alpha \hat a^{\dagger }-\alpha^{*}\hat a)=0$ holds, and the last term in the dissipation term cancels out with the driving term of strength $\varepsilon_{p}$ in the Hamiltonian. Then the master equation can be written as
\begin{align}
\frac{d}{dt} \rho_{ab}^{u_2}
&=-i\left[ \hat H_2, \rho_{ab}^{u_2} \right]+ \gamma_{a}\mathcal L[a]\rho_{ab}^{u_2}
\label{ms_m}
\end{align}
where Hamiltonian
\begin{align}
\hat H_2=&\Delta \hat{a}^{\dagger }\hat{a}+\sum_j\Big[\omega_{b_j}\hat{b}_j^{\dagger }\hat{b}_j+
g_{j}\hat{a}^{\dagger }\hat{a}(\hat{b}_j+\hat{b}_j^{\dagger})+(\alpha^{*}\hat{a}+\alpha \hat{a}^{\dagger}+|\alpha|^{2})(g_{j}\hat{b}_j+g_{j}\hat{b}_j^{\dagger})\Big]+(\varepsilon_{d}\hat{a}^{\dagger}e^{i\Delta_pt}
+h.c.).
\end{align}
It can be observed that the dissipation term still takes the form of single-photon dissipation.
The third and fourth parts in the Hamiltonian respectively characterize the nonlinear and  linearized optomechanical interactions. Generally, in the linearized cavity optomechanics, the nonlinear interactions are neglected for very weak single-photon optomechancal coupling, i.e., $r_j\equiv g_j/\omega_j\ll1$. In contrast, here we retain the nonlinear terms by assuming moderate single OM coupling (e.g., $r_j\sim0.1$).

Next we apply two successive Schrieffer-Wolff transformations, with the generator  $\hat S^{\dagger}=e^{r_2 \hat{a}^{\dagger }\hat{a}(\hat{b}_2^{\dagger }-\hat{b}_2)}e^{r_1\hat{a}^{\dagger }\hat{a}(\hat{b}_1^{\dagger }-\hat{b}_1)}$, to the master equation \ref{ms_m}  and the operator becomes
\begin{align}
&\hat S^{\dagger}\hat a^{\dagger}\hat S = \hat a^{\dagger}e^{r_1(\hat b_1^{\dagger }-\hat b_1)}e^{r_2(\hat b_2^{\dagger }-\hat b_2)}
=\hat {\widetilde{a}}^{\dagger}\nonumber\\
&\hat S^{\dagger}\hat b_j^{\dagger}\hat S = \hat b_j - r_2\hat a^{\dagger }\hat a\nonumber\\
&\hat S^{\dagger}\hat a^{\dagger }\hat a\hat S = \hat a^{\dagger }\hat a\nonumber\\
&\hat S^{\dagger}\hat b_j^{\dagger }\hat b_j\hat S = \omega_{b_j}(\hat b_j^{\dagger}-r_j \hat a^{\dagger}\hat a)(\hat b_j-r_j \hat a^{\dagger}\hat a)
=g_j r_j\hat a^{\dagger}\hat a\hat a^{\dagger}\hat a+\omega_{b_j}\hat b^{\dagger}\hat b_{j}-\hat a^{\dagger}\hat a(g_j \hat b_j+g_j \hat b_j^{\dagger })
\end{align}
The Hamiltonian after the Schrieffer-Wolff transformations $\hat H_3=\hat S^{\dagger}\hat H_2\hat S$  is obtained as:
\begin{align}
\hat H_3=&\widetilde{\Delta} \hat{a}^{\dagger}\hat{a}+\omega_{b_1}\hat{b}_1^{\dagger}\hat{b}_1+\omega_{b_2}\hat{b}_2^{\dagger}\hat{b}_2
+g_0
\hat{a}^{\dagger}\hat{a}\hat{a}^{\dagger}\hat{a}
+(\alpha^{*}\hat{ \widetilde{a}}+\alpha \hat{ \widetilde{a}}^{\dagger})(g_1 \hat{b}_1+g_1 \hat{b}_1^{\dagger}+g_{2}\hat{b}_2+g_{2}\hat{b}_2^{\dagger })\nonumber\nonumber\\
&+|\alpha|^2 (g_1 \hat{b}_1+g_1 \hat{b}_1^{\dagger}+g_{2}\hat{b}_2+g_{2}\hat{b}_2^{\dagger })
+2g_0\hat{a}^{\dagger}\hat{a}(\alpha^{*}\hat{\widetilde{a}}+\alpha \hat{\widetilde{a}}^{\dagger})+\varepsilon_{d}\hat{ \widetilde{a}}^{\dagger}e^{i\Delta_p t}+\varepsilon_{d}^{*}\hat{ \widetilde{a}}e^{-i\Delta_p t},
\end{align}
where the detuning $\widetilde{\Delta}=\left(\Delta+2|\alpha|^2g_0\right)\nonumber$ and the coupling $g_0 =- \left(r_1g_1+r_2g_2\right)$. The dissipation term becomes $\gamma_a \mathcal L[\hat {\widetilde{a}}]\hat \rho_{ab}^{u_3}$, where $\rho_{ab}^{u_3}=\hat S^{\dagger}\rho_{ab}^{u_2}\hat S$, and the dissipation operator is transformed into $\hat {\widetilde{a}}$. We will discuss this further later on.

The operator $\hat {\widetilde{a}}^{\dagger}$ can be approximated, for $r_{j}\gg r_{j} r_{j'}$, as
\begin{align}
\hat{ \widetilde{a}}^{\dagger}
=& \hat{a}^{\dagger }\big[1+\frac{r_1}{1!}(\hat{b}_1^{\dagger}-\hat{b}_1)+\frac{(r_1)^2}{2!}(\hat{b}_1^{\dagger}-\hat{b}_1)^{2}+\cdots \big]\times\big[1+\frac{r_2}{1!}(\hat{b}_2^{\dagger}-\hat{b}_2)+\frac{(r_2)^2}{2!}(\hat{b}_2^{+}-\hat{b}_2)^{2}+\cdots \big]\nonumber\\
\approx& \hat{a}^{\dagger}\big[1+r_1(\hat{b}_1^{\dagger}-\hat{b}_1)
+r_2(\hat{b}_2^{\dagger}-\hat{b}_2)\big].
\label{efaA}
\end{align}
Substituting Eq.(\ref{efaA}) into the Hamiltonian $\hat H_3$ and applying the unitary transformation $e^{i\hat{\widetilde{ H_0}}t}$, where $
\hat{\widetilde{H_0}}=\widetilde{\Delta}\hat{a}^{\dagger}\hat{a}+\omega_{b_1}\hat{b}_1^{\dagger}\hat{b}_1+\omega_{b_2}\hat{b}_2^{\dagger}\hat{b}_2 $, the Hamiltonian becomes
\begin{align}
\hat H_4 =& e^{i\hat{\widetilde{ H_0}}t}   [g_0\hat{a}^{\dagger}\hat{a}\hat{a}^{\dagger}\hat{a}
+\alpha \hat{ \widetilde{a}}^{\dagger}(g_1 \hat{b}_1+g_1 \hat{b}_1^{\dagger}+g_{2}\hat{b}_2+g_{2}\hat{b}_2^{\dagger }
+2g_0\hat{a}^{\dagger}\hat{a} +\frac{\varepsilon_{d}}{\alpha}e^{i\Delta_p t})\nonumber\\
&+|\alpha|^2 (g_1 \hat{b}_1^{\dagger}+g_{2}\hat{b}_2^{\dagger })
+h.c.  ] e^{-i\hat{\widetilde{ H_0}}t}\nonumber\\
\approx &e^{i\hat{\widetilde{ H_0}}t}   \Big \{g_0\hat{a}^{\dagger}\hat{a}\hat{a}^{\dagger}\hat{a}
+|\alpha|^2 (g_1 \hat{b}_1^{\dagger}+g_{2}\hat{b}_2^{\dagger })+\alpha \hat{a}^{\dagger}\big[1+r_1(\hat{b}_1^{\dagger}-\hat{b}_1)
+r_2(\hat{b}_2^{\dagger}-\hat{b}_2)\big]\nonumber\\
&\times (g_1 \hat{b}_1+g_1 \hat{b}_1^{\dagger}+g_{2}\hat{b}_2+g_{2}\hat{b}_2^{\dagger }
+2g_0\hat{a}^{\dagger}\hat{a} +\frac{\varepsilon_{d}}{\alpha}e^{i\Delta_p t})
+h.c.  \Big\} e^{-i\hat{\widetilde{ H_0}}t}\nonumber\\
=&g_0\hat a^{\dagger}\hat a\hat a^{\dagger}\hat a
+\alpha\hat a^{\dagger}e^{i\widetilde{\Delta} t}\left[1+r_1(\hat b_1^{\dagger}e^{i\omega_{b_1}t}-\hat b_1e^{-i\omega_{b_1}t})+r_2(\hat b_2^{\dagger }e^{i\omega_{b_2}t}-\hat b_2e^{-i\omega_{b_2}t})\right] \nonumber\\
&\times\left[g_{1}\hat b_1e^{-i\omega_{b_1}t}+g_{1}\hat b_1^{\dagger }e^{i\omega_{b_1}t}+g_{2}\hat b_2e^{-i\omega_{b_2}t}+g_{2}\hat b_2^{\dagger }e^{i\omega_{b_2}t}+2g_0\hat{a}^{\dagger}\hat{a}+\frac{\varepsilon_{d}}{\alpha}e^{i\Delta_p t} \right]\nonumber\\
&+|\alpha|^2(g_{1}\hat b_1^{\dagger }e^{-i\omega_{b_1}t}+g_{2}\hat b_2^{\dagger }e^{-i\omega_{b_2}t})+h.c.
\end{align}
The Hamiltonian can be transformed under the condition $\widetilde{\Delta}=-\Delta_p=\omega_{b_1}+\omega_{b_2}$ (i.e., Eq.\ref{reltA}) into
\begin{align}
\hat H_4
\approx g_0
\hat{a}^{\dagger}\hat{a}\hat{a}^{\dagger}\hat{a}
+g\hat{a}^{\dagger}\hat{b}_1 \hat{b}_2 + g^{*}\hat{a} \hat{b}_1^{\dagger} \hat{b}_2^{\dagger} + \varepsilon_{d}\hat{a}^{\dagger} + \varepsilon_{d}^{*}\hat{a},
\label{h_aA}
\end{align}
where $g = -\alpha \left(r_1g_2+r_2g_1\right)$. The above Hamiltonian is obtained under the rotating wave approximation requiring the following conditions
\begin{align}
\{\omega_{b_j},|\omega_{b_1}-\omega_{b_2}|,\widetilde{\Delta},
|\widetilde{\Delta}-\omega_{b_j}|\}\gg
\{r_j\varepsilon_{d},g_j|\alpha|^2,
 &\alpha \left(r_1g_2+r_2g_1\right),
\alpha \left(r_1g_1+r_2g_2\right)\}.
\label{eq_con}
\end{align}

In Eq.(\ref{h_aA}), the first term is the effective cavity Kerr nonlinearity, the second (third) term describes the hybrid four-wave mixing process among the cavity photons, strong-drive photons and phonons, in which a drive photon and two nondegenerate phonons are simultaneously annihilated and then a cavity photon ($\omega_c\approx\omega_p+\sum_j\omega_{b_j}$) is created (vice versa). This process can lead to non-Gaussian quantum correlations between the two mechanical oscillators, under the cavity driving on resonance described by the last two terms.
Next, we investigate the dissipation term under the unitary transformation.
\begin{align}
e^{i\hat{\widetilde{ H_0}}t}&\mathcal L[\hat { \widetilde{a}}]\hat \rho_{ab}^{u_3}e^{-i\hat{\widetilde{ H_0}}t}
=e^{i\hat{\widetilde{ H_0}}t}\left(2\hat{ \widetilde{a}}\hat \rho_{ab}^{u_3} \hat{ \widetilde{a}}^{\dagger}-\hat a^{\dagger}\hat a\hat \rho_{ab}^{u_3}-\widetilde{\rho } \hat a^{\dagger}\hat a\right)e^{-i\hat{\widetilde{ H_0}}t}\nonumber\\
\approx& -\hat a^{\dagger}\hat a\hat \rho_{ab}^{u_4}-\hat \rho_{ab}^{u_4}\hat a^{\dagger}\hat a \nonumber\\
&+2e^{i\hat{\widetilde{ H_0}}t}\hat a\left[1+r_1(\hat b_{1}-\hat b_{1}^{\dagger})
+r_2(\hat b_{2}-\hat b_{2}^{\dagger})\right]\hat \rho_{ab}^{u_3}a^{\dagger}
\left[1+r_1(\hat b_{1}^{\dagger}-\hat b_{1})+r_2(\hat b_{2}^{\dagger}-\hat b_{2})\right]
e^{-i\hat{\widetilde{ H_0}}t}\nonumber\\
=& -\hat a^{\dagger}\hat a\hat \rho_{ab}^{u_4}-\hat \rho_{ab}^{u_4}\hat a^{\dagger}\hat a \\
&+2\hat a\left[1+r_1(\hat b_{1}e^{-i\omega_{b_1}t}-\hat b_{1}^{\dagger}e^{i\omega_{b_1}t})
+r_2(\hat b_{2}e^{-i\omega_{b_2}t}-\hat b_{2}^{\dagger}e^{i\omega_{b_2}t})\right]\nonumber\\
&\times \hat \rho_{ab}^{u_4}a^{\dagger}
\left[1+r_1(\hat b_{1}^{\dagger}e^{i\omega_{b_1}t}-\hat b_{1}e^{-i\omega_{b_1}t})
+r_2(\hat b_{2}^{\dagger}e^{i\omega_{b_2}t}-\hat b_{2}e^{-i\omega_{b_2}t})\right]\nonumber\\
\end{align}
Under condition (\ref{eq_con}), the rotating wave approximation is applied, and considering condition $r_{j}\gg r_{j} r_{j'}$, the dissipation term can be approximated as:
\begin{align}
e^{i\hat{\widetilde{ H_0}}t}\mathcal L[\hat { \widetilde{a}}]\hat \rho_{ab}^{u_3}e^{-i\hat{\widetilde{ H_0}}t}
\approx&2\hat a\hat \rho_{ab}^{u_4}\hat a^{\dagger}-\hat a^{\dagger}\hat a\hat \rho_{ab}^{u_4}
-\hat \rho_{ab}^{u_4}\hat a^{\dagger}\hat a\nonumber\\
&+2\left[r_1^2(\hat a\hat b_1\rho_{ab}^{u_3}\hat a^{\dagger}\hat b_1^{\dagger}
+\hat a^{\dagger}\hat b_1\rho_{ab}^{u_3}\hat a^{\dagger}\hat b_1)+r_2^2(\hat a\hat b_2\rho_{ab}^{u_3}\hat a^{\dagger}\hat b_2^{\dagger}+\hat a^{\dagger}\hat b_2\rho_{ab}^{u_3}\hat a^{\dagger}\hat b_2)\right]\nonumber\\
\approx& 2\hat a\hat \rho_{ab}^{u_4}\hat a^{\dagger}
-\hat a^{\dagger}\hat a\hat \rho_{ab}^{u_4}-\hat \rho_{ab}^{u_4}\hat a^{\dagger}\hat a
= \mathcal L[\hat a] \hat \rho_{ab}^{u_4}.
\end{align}

Here, the dissipation operator is $\hat a$, and under the approximate conditions, it still takes the form of single-photon dissipation. In the transformed picture, the master equation (\ref{ms_bA}) is approximately reduced to
\begin{align}
\frac{d}{dt} \hat \rho_{ab}^{u_4}
&\approx -i\left[ \hat H_4 , \hat\rho_{ab}^{u_4} \right]+ \gamma_{a}\mathcal L[\hat{a}]\hat\rho_{ab}^{u_4}\nonumber\\
\label{ms_bA}
\end{align}
It can be inferred from the above master equation that, in the absence of the mechanical damping ($\gamma_{b_j}=0$), the cavity and mechanical system is dissipated by the cavity dissipation into a steady dark state, i.e., $\hat \rho_{ab}^{\rm ss}=|\zeta\rangle_b|0\rangle_a \langle 0|_b\langle \zeta|$, where the cavity field is in vacuum and the mechanical state $\left | \zeta \right \rangle_b$  satisfies the equation $(g\hat{b}_1\hat{b}_2+\varepsilon _d)\left | \zeta \right \rangle_b=0$. This shows that the mechanical oscillators are dissipatively driven into the eigenstate of the operator $\hat{b}_1\hat{b}_2$, i.e.,
\begin{align}
 \hat{b}_1\hat{b}_2\left | \zeta \right \rangle_b=\zeta \left | \zeta \right \rangle_b,
 \end{align}
where the eigenvalue $\zeta = -\varepsilon _d/g$. For simplicity, we assume that the two mechanical oscillators the initial phonon number difference $\hat b_1^{\dagger}\hat b_1-\hat b_2^{\dagger}\hat b_2=0$, then the eigenvalue equation becomes
\begin{align}
 \hat{b}_1\hat{b}_2\left | \zeta \right \rangle_b&=\zeta \left | \zeta \right \rangle_b\nonumber\\
\hat{b}_1\hat{b}_2\sum_{m=0 }^{\infty} C_{m} \left | m \right \rangle_{b_1}\left | m \right \rangle_{b_2}
&=\zeta\sum_{n=0 }^{\infty} C_{n} \left | n \right \rangle_{b_1}\left | n \right \rangle_{b_2}\nonumber\\
\sum_{k=0 }^{\infty} C_{k+1}(k+1) \left | k \right \rangle_{b_1}\left | k \right \rangle_{b_2}
&=\zeta\sum_{n=0 }^{\infty} C_{n} \left | n \right \rangle_{b_1}\left | n \right \rangle_{b_2}\nonumber\\
\Rightarrow C_{n+1}=\frac{\zeta }{n+1} C_{n}&\Rightarrow C_{n}=\frac{\zeta^n }{n!}
 \end{align}
The steady mechanical state of master equation (\ref{ms_bA}) in Fock space is obtained as
\begin{align}
\left | \zeta \right \rangle_b  \propto \frac{1}{\mathcal N_{0} }\sum_{n=0 }^{\infty }
\frac{\zeta ^{n} }{n! }   |  n   \rangle_1| n \rangle_2,
\end{align}
This state is called a PCS, which is analogous to a coherent state as the eigenstate of a bosonic annihilation operator. Similar to a two-mode squeezed vacuum, the PCS is also a superposition only of states in which the two modes contain the same number of phonons. However,  the probability amplitudes of the PCS differ from those of the two-mode squeezed vacuum, which has Gaussian-type quasiprobability distribution in phase space. As a result, the PCS is a non-Gaussian state.
It is noteworthy that $\hat \rho_{ab}$ and $\hat \rho_{ab}^{u_4}$ represent the same state in different representations. Therefore, when calculating the fidelity between the solution $\hat \rho_{ab}$ of the master equation (\ref{ms_oA})  and the PCS, it is necessary to use the unitary transformation $\hat U^{\dagger}=e^{i\widetilde{H_0}t}e^{r_2 \hat{a}^{\dagger }\hat{a}(\hat{b}_2^{\dagger }-\hat{b}_2)}e^{r_1\hat{a}^{\dagger }\hat{a}(\hat{b}_1^{\dagger }-\hat{b}_1)}e^{\alpha^{*}\hat{a}-\alpha \hat{a}^{\dagger}}e^{i\omega_{p}\hat{a}^{\dagger}\hat{a}t}$] to convert $\hat \rho_{ab}$ into the representation of $\hat \rho_{ab}^{u_4}$.

\end{widetext}
\providecommand{\noopsort}[1]{}\providecommand{\singleletter}[1]{#1}%

\end{document}